\documentclass[12pt, draftclsnofoot, onecolumn]{IEEEtran}

\usepackage{verbatim}
\usepackage{xpatch}
\usepackage{subfigure}
\usepackage{graphicx}
\usepackage[ruled, linesnumbered]{algorithm2e}
\usepackage{amssymb}
\usepackage{amsmath} 
\usepackage{caption}
\usepackage{xcolor}
\usepackage{soul}

\ifCLASSOPTIONcompsoc
\usepackage[nocompress]{cite}
\else
\usepackage{cite}
\fi

\newcommand{\biaoti}{\fontsize{23.4pt}{29pt}\selectfont}

\begin{document}
\title{\biaoti Multi-agent Reinforcement Learning for Dynamic Resource Management in 6G in-X Subnetworks}

	\author{Xiao~Du,~\IEEEmembership{}
	Ting~Wang,~\IEEEmembership{Senior Member, IEEE,}
	Qiang~Feng,~\IEEEmembership{}
	Chenhui~Ye,~\IEEEmembership{}
	Tao~Tao,~\IEEEmembership{} \\
	Yuanming Shi,~\IEEEmembership{Senior Member, IEEE}
	and Mingsong Chen,~\IEEEmembership{Senior Member, IEEE}
	\thanks{ X. Du, T. Wang and M. Chen are with the Shanghai Key Laboratory of Trustworthy Computing, East China Normal University, Shanghai 200062, China (e-mail: 71194501039@stu.ecnu.edu.cn, twang@sei.ecnu.edu.cn, mschen@sei.ecnu.edu.cn). X. Du is also with the Bell Labs, Shanghai, China. Q. Feng, C. Ye, T. Tao are with the Bell Labs, Nokia Shanghai Bell Corp., Shanghai 201206, China (email:qiang.feng, chenhui.a.ye, tao.b.tao@nokia-sbell.com). Y. Shi is with the School of Information Science and Technology, ShanghaiTech University, Shanghai 201210, China (e-mail: shiym@shanghaitech.edu.cn).}
}



\maketitle

\begin{abstract}

The 6G network enables a subnetwork-wide evolution, resulting in a “network of subnetworks". However, due to the dynamic mobility of wireless subnetworks, the data transmission of intra-subnetwork and inter-subnetwork will inevitably interfere with each other, which poses a great challenge to radio resource management. Moreover, most of the existing approaches require the instantaneous channel gain between subnetworks, which are usually difficult to be collected. To tackle these issues, in this paper we propose a novel effective intelligent radio resource management method using multi-agent deep reinforcement learning (MARL), which only needs the sum of received power, named received signal strength indicator (RSSI), on each channel instead of channel gains. However, to directly separate individual interference from RSSI is an almost impossible thing. To this end, we further propose a novel MARL architecture, named GA-Net, which integrates a hard attention layer to model the importance distribution of inter-subnetwork relationships based on RSSI and exclude the impact of unrelated subnetworks, and employs a graph attention network with a multi-head attention layer to exact the features and calculate their weights that will impact individual throughput. Experimental results prove that our proposed framework significantly outperforms both traditional and MARL-based methods in various aspects.


\end{abstract}


\section{Introduction} \label{sec:introduction}
The cellular systems of 2G/3G/4G are designed primarily for the voice and data service. The fifth-generation (5G) mobile communication system  is the first system designed to make inroads into the industrial environment. This will extend the carriers' and vendors' business scope to vertical markets, and change their business mode to increase the income. 
Recently, the emerging sixth-generation (6G) technology enables various new revolutionary services, for example, high-resolution sensing and pervasive mixed reality, requiring extreme performance in terms of latency (down to 100 µs), reliability (for life-critical applications), and throughput (Gbit/s for AR/VR). 
Owing to the higher reliability, the lower latency, and the increased data rate in resource allocation of wireless systems, industrial wireless networks are expected to supersede the cumbersome traditional wired industrial network infrastructure like EtherCAT (Ethernet for Control Automation Technology), Profinet, and the time sensitive network (TSN) solutions. According to European 6G white paper  \cite{dixon_2017}, the increasing number of end devices will evolve in a variety of scenarios to be a network of devices, also known as a subnetwork. Independent and uncoordinated subnetworks have been identified as a promising solution for supporting extreme connectivity in recent visions on 6G \cite{letaief2019roadmap1}. The concepts and design principles for such 6G in-X subnetworks are exhibited in \cite{dixon_2017}. To further emphasize the term in-X for inside-everything, the work \cite{20206G4} clarifies a series of emerging in-X scenes, including in-robots, in-airplanes, in-vehicles, and in-human bodies. The connectivity scenarios are various, including static and isolated devices, as well as interconnected local interactive devices and fast moving drones or robots, which connect to a common cellular network.

\begin{figure}[htb]  
	\centering  
	\includegraphics[width=16cm]{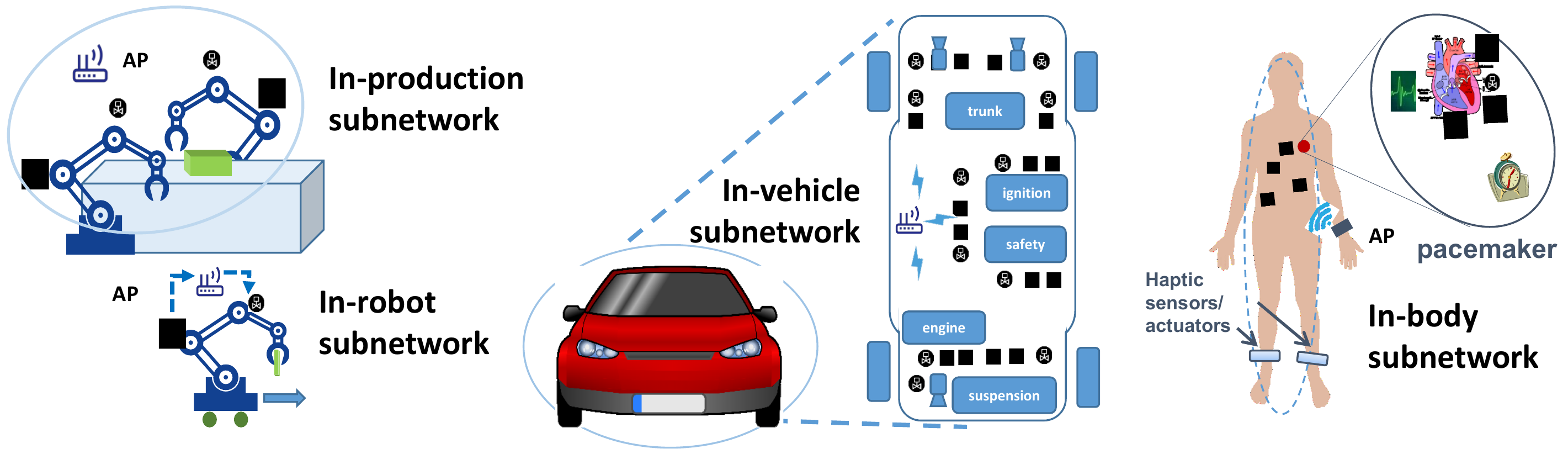}  
	\caption{An example of 6G in-X subnetworks.} \label{framework}
\end{figure}

Fig. \ref{framework} depicts several different kinds of in-X subnetworks, including in-production subnetwork, in-vehicle subnetwork, in-robot subnetwork and in-body subnetwork. In addition, Fig. \ref{subnetworks} gives an example of subnetworks, each of which connects multiple functional parts, including a controller, some sensors to gather machine status, and some actuators responsible for operating the machines. The communications between controller-sensors and controller-actuators are wireless, which share the same frequency bands with cellulars. 
To ensure high reliability and determinism in both the temporal and spatial domain, the subnetworks will remain uninterrupted despite poor or no connection to the wider networks. However, the rapid moveability of the subnetwork would potentially trigger highly dynamic interference, which will result in intolerable high transmission failure rates. 
In order to mitigate such high interference power, effective radio resource allocation algorithms should be adopted to maximize the utilization of available multi-dimensional radio resources (such as frequency band resources and transmission power budgets) under dynamic interference conditions with ultra-tight delay constraints. The radio resource management problem is usually non-convex with NP-hardness, lacking effective combinatorial universal solutions \cite{2021Radio}.

To tackle such computationally intractable problems, there have been many approaches, leveraging techniques in various fields, for example, geometric programming \cite{46002287}, weighted minimum mean square optimization \cite{57564896}, game theory \cite{68450588}\cite{matching2015}\cite{spectrum2017}, fractional programming \cite{2017FPLinQ9}, information theory \cite{682474510}\cite{2015ITLinQ11}  and machine learning \cite{875530012}\cite{2017Artificial13}. 
However, in order to enable dynamic resource allocation optimization, these existing algorithms, no matter conventional Centralized Graph Coloring (CGC) algorithm \cite{2020Distributed26} or machine learning-based methods, typically depend on sorts of hardly accessible information in a real-world network, such as channel gain between any two subnetworks as there are no direct communications between them in practice. 
In addition, the existing methods are difficult to reason the potential interference relationships between agents in multiple mobile subnetwork systems. The potential interference relationship can be understood as the probability of continuously selecting the same channel and the distance between subnets, etc.

Recently, deep reinforcement learning (DRL) has demonstrated its superiority with excellent performance in solving the problem of radio resource allocation \cite{2019Spectrum15}\cite{li2019multi}.
Although a variety of excellent approaches have been proposed, there still exist some intractable issues in these works. On the one hand, the existing methods require relying on instantaneous information, which is difficult to obtain, such as the instantaneous channel gain between subnetworks. On the other hand, as the number of subnetwork grows, the computing complexity of training will increase dramatically.

Motivated by these observations, in this paper, we propose a novel framework for dynamic resource allocation in 6G in-X subnetworks based on multi-agent deep reinforcement learning (MARL), where each subnetwork is treated as an agent that automatically learns to refine a reasonable resource management policy for transmission. Notably, our proposed framework, which shares historical states and actions of all subnetworks during centralized training, executes in a distributed manner and trains in a centralized manner.
Moreover, our proposed DRL-based framework uses RSSI as its input, which is composed of the intra-subnetwork transmission power and all the inter-subnetwork interference, instead of instantaneous channel state information.

However, due to the continuously moving environment, variable transmitting power levels and time varying fast fading, the interference is not constant, so it is nearly impossible to directly separate the individual interference from RSSI.  In this context, our approach models the subnetwork system as a complete graph and employs a graph neural network (GNN) combining with two-stage attention networks to effectively reason the inter-subnetwork relationships. 
Specifically, our method utilizes an improved hard attention to eliminate the impact of the unrelated subnetworks, which is conducive to decreasing the computing complexity and simplifying the relationship among subnetworks.
Then, GNN represents the subnetwork system as a time-varying graph, where some edges with weak correlation will be cut off by the hard attention. Multi-head attention is used to decouple and reason various potential interference relations among subnetworks from RSSI. At last, each subnetwork learns the policy to coordinate resource allocation. Our approach shares historical and current state information of all subnetworks instead of only current state information, which is beneficial to reasoning potential interference relationship among subnetworks from historical information during centralized training. It is worth noting that historical information is not used to make decisions, but only to help train policies that can make more reasonable decisions.

The key contributions of this paper are summarized s follows:
\begin{itemize}
	\item We propose a novel resource management framework for 6G in-X multi-subnetworks based on MARL, which can effectively extract the inter-subnetwork interference relationships from RSSI.
	\item We propose a new soft actor-critic based training algorithm, which uses RSSI at each spectrum band as the state input to MARL, without requiring any prior knowledge about the hardly accessible information such as source output power and the channel gains.	
	\item We propose a novel graph attention network, named GA-Net, by combining a two-stage attention network and graph neural network, to decouple and reason the potential interference relations among subnetworks from RSSI, which is beneficial to learn an intelligent policy to cooperatively select channels and power levels, and meanwhile decrease the computational complexity during centralized training.
	\item We conduct extensive experiments to show the effectiveness and efficiency of our approach. The experimental results prove that our approach outperforms the existing schemes.
\end{itemize}

The rest of this paper is organized as follows. Section \ref{Related_work} briefly reviews the related work.  Section \ref{BACKGROUND} and \ref{system model} present the preliminary knowledge and system model design, respectively.  The resource allocation problem is formulated as the MARL model in Section \ref{RL model}. Section \ref{approach} details the design of our proposed approach. 
Section \ref{evaluation} presents performance evaluation results. Finally, the whole paper concludes in Section \ref{conclusion}.

\section{Related Work}\label{Related_work}

In this section, the prior works on radio resource allocation are briefly investigated and summarized. Generally, the existing approaches for radio resource allocation can be categorized into two groups: centralized schemes and distributed schemes. 

\subsection{Centralized Schemes}
Among the centralized approaches, the work \cite{2020Distributed26} lists some possible algorithms for subnetwork resource allocation, including the minimum SINR (signal to interference-plus-noise ratio) guarantee algorithm, the Nearest Neighbour Conflict Avoidance (NNAC) algorithm and the CGC algorithm. 
All these are centralized algorithms, on top of the issue that they can’t access the unavailable channel gains between subnetworks, they also generate massive data traffic due to huge data exchange during the iterative resource allocation optimization. 
In work \cite{resource2013}\cite{resource2018}, the cellular users (CUEs) rely on the cellular base station (BS) to allocate resources, and monitor information like the SINR and the channel state information (CSI). By leveraging the global CSI at BS, the work \cite{learning2013} proposes an effective wireless resource allocation algorithm based on graph theory. The work \cite{9252917} presents a graph neural network based approach to deal with the large-scale radio resource management issues. However, such centralized schemes have a major limitation, that is, the global network information is required. Worse still, the computing complexity of these centralized approaches dramatically rises as the number of subnetworks grows, resulting in an immense computational burden on base stations. The authors in \cite{2019A20} apply a centralized deep Q-learning method to achieve downlink power control, where the agent can obtain the global network state and make power control decisions for all transmitters.

\subsection{Distributed Schemes}
Aiming to decrease the signaling overhead and the computing load, many distributed resource allocation methods have been proposed, without any central controllers \cite{732222617}\cite{2016A18}. In work \cite{2017Distributed16}, game theory is applied to model device-to-device (D2D) pairs, and an auction-based algorithm is proposed to achieve spectrum resource sharing in a distributed way.
However, this algorithm converges slowly requiring a large number of iterations, and users need to exchange channel gain information with each other. 
DRL methods have shown significant potentials in resource allocation in recent studies. The authors in \cite{2016A18} consider a cellular network where users in each cell get an equal share of the spectrum, and use deep MARL to complete power control and rate adaptation to optimize network-wide utility function. The work \cite{876143119} applies deep Q-learning to solve the power control problem aiming to maximize the averaged sum-rate in multi-user cellular networks. 
In \cite{879211721}, the authors design a distributed downlink power control scheme based on deep MARL, where each transmitter collects observations from its neighbouring transmitters at each scheduling interval and then makes power control decisions to maximize the weighted sum-rate. 
The work \cite{9565875} proposes a small base stations state selection scheme based on multi-agent deep reinforcement learning to solve the joint optimization problem of massive access and resource management in Ultra-dense network where human type communications and machine type communications coexist.
The work \cite{lu2021dynamic} proposes a centralized training reinforcement learning method DRL-CT to solve the problem of joint resource allocation. In addition, a federated deep reinforcement learning algorithm which can reduce communication overhead and protect user privacy is proposed to imitate DRL-CT. The work \cite{naderializadeh2021resource} proposes a multi-agent deep reinforcement learning algorithm for distributed resource management and interference mitigation in wireless networks. In this algorithm, the observation and action space of agents is scalable, so that the policies trained can be migrated to the scene with different number of agents.
The work \cite{2021Distributed} proposes two distributed algorithms based on single agent reinforcement learning, which is suitable for the media access scenario based on competition. The algorithm can let the base station choose the best transmission modulation scheme in each time slot, so as to maximize the proportional fairness of UE throughput.

\section{Preliminaries}\label{BACKGROUND}
In this section, some preliminary background knowledge about our proposed MARL-based framework is introduced.

\subsection{Multi-Agent Reinforcement Learning}
With the burgeoning of reinforcement learning (RL) and deep learning (DL), RL research has shifted from a single agent to a more challenging and practical multi-agent. Existing MARL studies in this area have primarily concentrated on deriving decentralized policies, which are trained with the Centralized Training and Decentralized Execution (CTDE) framework. These methods can be divided into two groups, namely the learning-for-consensus scheme and the learning-to-communicate scheme, depending on how the consensus among multiple agents is derived. For example, MADDPG \cite{2017Multi33} is a learning-for-consensus approach, which employs the CTDE framework and derives decentralized policies for competitive or cooperative tasks. TarMAC \cite{tarmac2019} is a learning-to-communicate approach, which uses an attention network based to learn communication protocol to make individual decisions. In work \cite{liu2020multi}, the attention mechanism is introduced to solve the problem that the learning complexity increases 
rapidly with the increase of the number of agents.

The work \cite{2018Value31} first proposes VDN, which is a method based on value function decomposition. However, it simply believes that the joint Q-value function is the simple addition of local Q-value functions of other agents. QMIX \cite{2018QMIX32} improves VDN, where the joint Q-value function is estimated as a complex nonlinear combination of the local Q-value function of the agents. Moreover, it is emphasized that the joint Q-value is monotonic in the local Q-value function. The work \cite{son2019qtran} is a further improvement of the VDN and QMIX algorithms, which first uses the VDN method to obtain the summed local Q-value function as an approximation of the joint Q-function, and then fits the difference between the local Q-function and the joint Q-function. The work \cite{wang2020qplex} adopts a duplex dueling network architecture to decompose the joint value function, which makes the q-value function of a single agent approach the global maximization.

The work \cite{iqbal2021randomized} proposes a algorithm, which randomly divides agents into two groups, and counterfactually reasons the utilities of the agent groups based on historical information, and then uses these imagined utility experiences to improve the prediction of utility functions in the real environment. The work \cite{jiang2018graph} utilizes self attention to obtain convolution kernel, and then expands the receptive field of agents through multi-layer convolution, so as to obtain a information-condensed state representation , which is conducive to learning the abstract relationship between agents. The work \cite{naderializadeh2020graph} proposes an approach, which incorporates graph neural networks into a multi-agent reinforcement learning approach based on value function decomposition.
The work \cite{malysheva2019magnet} proposes a method, which utilizes the self attention mechanism to learn the relevance graph, and then uses it together with the state information for reinforcement learning training.  The work \cite{bai2021value} proposes a multi-agent reinforcement learning algorithm that combines  hypergrap convolution and value function decomposition, which explores the relationship between action values by self-learning hypergraphs. 

\subsection{Soft Actor-critic}
Soft Actor-Critic (SAC) \cite{2018Soft34} is designed to train stochastic policies, and it introduces the approach of Maximum Entropy Reinforcement Learning (MERL) to learn a soft value estimate by incorporating an entropy term in the learning value function, i.e.,
\begin{align}
J(\pi) = \mathbb{E}_{\tau \sim \pi} \left[ \sum_{t=0}^{\infty} \gamma ^ t r(s_t,a_t) + \alpha H(\pi(\cdot\lvert s_t)) \right], 
\end{align}
where  $H(\pi(\cdot\lvert s_t))$ indicates the entropy of the policy,  $\pi$ is a policy for mapping observation to action distribution, $\gamma$ is a discount factor, $\tau$ denotes a trajectory of consecutive states and actions and $\alpha$ denotes the temperature parameter indicating the relative importance of the reward and the entropy. Besides, the AC algorithm's target $Q_{targ}$ and policy gradient $\nabla_{\theta}J$ are respectively tuned as
\begin{align}
Q_{targ}(s_t, a_t) &= r(s_t, a_t) + \gamma\mathbb{E}_{ a_{t+1} \sim \pi_{\hat{\theta}}(s_{t+1})} [ Q_{\hat{\psi}}(s_{t+1}, a_{t+1}) -\alpha log(\pi_{\hat{\theta}}(a_{t+1} \lvert s_{t+1})) ], &
\end{align} 
\begin{align}
\nabla_{\theta}J(\pi_{\theta}) &= \mathbb{E}_{s_t \sim D, a_t \sim \pi_\theta}[\nabla_{\theta}log(\pi_{\theta}(a_t \lvert s_t))(\alpha log(\pi_{\theta}(a_t \lvert s_t)) - Q_{\psi}(s_t,a_t) + b(s_t))], & 
\end{align}
where $Q_{\hat{\psi}}$ and $\pi_{\hat{\theta}}$ are respectively the target Q-value function and the target policy function, $Q_{\psi}$ and $\pi_{\theta}$ are respectively the Q-value function and the policy function, and $D$ is the replay buffer. $b(s_t)$ is a common trick introduced in policy gradient reinforcement learning to reduce the variance in the learning process, and it is generally equal to the Q-value function in this state.  The objective of entropy regularization in SAC algorithm is to maximize the tradeoff between cumulative rewards and the entropy of policy distribution. Each agent is encouraged to avoid converging to a locally optimal solution in the stage of training.  To solve the problem of resource allocation, we extend the SAC algorithm to the multi-subnetwork environment and extended to discrete action space to encourage the subnetworks to find the optimal channels and powers selection behavior during the training phase, thus leading to obtain policies that can better cooperate with other subnetworks to maximize global resource utilization.

\subsection{Soft Attention and Hard Attention}
Hard attention and soft attention mechanisms are widely applied in machine learning systems, for example, natural language processing and computer vision. 
The hard attention mechanism forces a model to only calculate the weights of important elements, while completely discarding other elements. However, the hard attention mechanism cannot be trained through end-to-end backpropagation due to its selecting elements based on sampling. Therefore, we adopt the Gumbel Softmax estimator \cite{2016Categorical35} to calculate the weights. In this paper, we employ an improved hard attention to exclude other subnetworks with little correlation and reduces the complexity of the relationship between subnetworks, which decreases the difficulty of reasoning the interaction relation between subnetworks and computational complexity during training.
Soft attention calculates the importance distribution of various influencing elements. In particular, the soft attention is completely differentiable, so it can be easily trained through end-to-end backpropagation, where the softmax function is a commonly used activation function. We utilize multi-head attention, which can jointly attend to information from different representation subspaces, to decouple and reason various potential interference relations among subnetworks from RSSI.

\subsection{Graph Neural Networks}
Graph Neural Network (GNN) is a novel neural network architecture, which can derive the dependent relationships among nodes in the graph via message dissemination among graph nodes. 
Therefore, GNN can effectively address the learning problem using a graphic architecture. One of the most representative graph neural networks is Message Passing Neural Network (MPNN), in which each vertex integrates the feature information received from adjacent vertex to obtain the hidden state embedding depending on graph perception. 
In practical terms, the hidden state embedding of each node is updated iteratively through gathering state information from its adjacent nodes. In this paper, GNN is used to represent the subnetwork system as a time-varying graph, and then a two-stage attention mechanism is utilized to simplify the graph and extract the potential interference relationship among subnetworks from multiple dimensions.

To sum up, currently, MARL algorithms with attention mechanisms and GNN have been gradually adopted to deal with the resource management problems. 
Instead of directly using the existing MARL algorithm, in order to maximize the efficiency of radio resource management, we propose a novel MARL algorithm with some purposeful improvements, which are mainly reflected in the following three aspects. Firstly, during centralized training, our method utilizes GRU to fuse the current observation and historical information of the agent to obtain an information-condensed state representation as critic input, which provides rich information for inferring potential interference between agents. Secondly, our method combines GRU and Gumbel softmax estimator to form a hard attention module, which can generate the weight of $0$ or $1$ and realize the back propagation of the gradient at the same time, so as to overcome the limitation that hard attention can not obtain the gradient. Thirdly, our method integrates the multi head attention mechanism in GAT to form multiple subgraphs with different attention weights for each agent, so that our method can infer a variety of potential interference relationships.

\section{Subnetwork System Model} \label{system model}
\begin{figure}[htb]  
	\centering  
	\includegraphics[width=13.5cm]{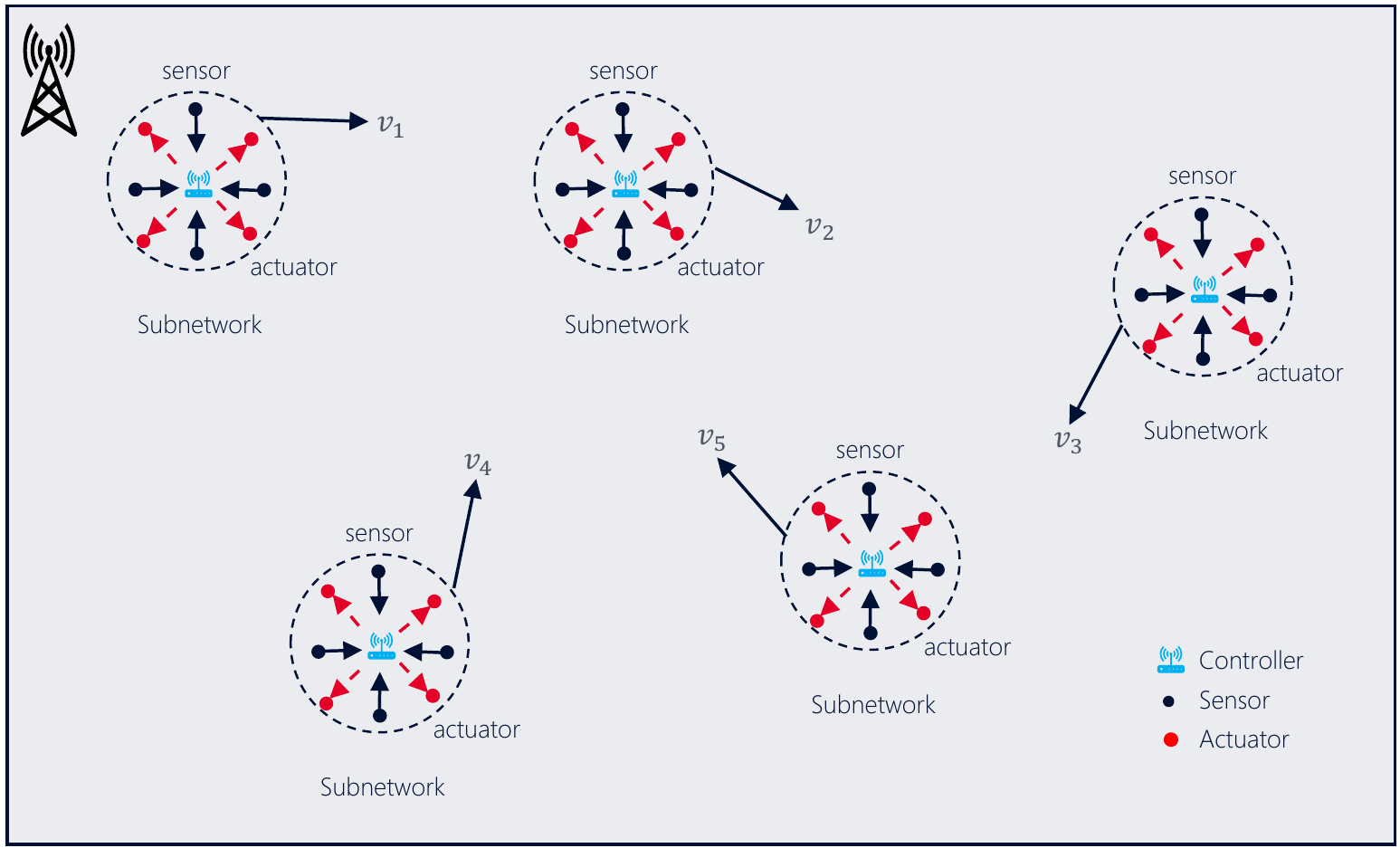}  
	\caption{An example of subnetworks.} \label{subnetworks}
\end{figure}

This section elaborates the model of subnetworks, where a variety of connectivity scenarios are considered, including subnetwork of static devices, fast moving devices (e.g. drones), isolated devices, and local interacting devices. 
Within a subnetwork, the orthogonality of communication links should be guaranteed to ensure extreme communication performance. However, multiple subnetworks share the same frequency ranges, resulting in intense competition for resources. Suppose that there are $N$ subnetworks, and each subnetwork $i \in \{1,2,...,N\}$ contains a controller (i.e., AP), $K$ sensors, and $K$ actuators. 
The controller is expected to regularly gather data from the sensors, and control the actuators periodically. The total bandwidth is partitioned into $M$ channels, and each subnetwork needs a channel for intra-subnetwork communications. Fig. \ref{subnetworks} depicts an example of subnetworks.

Considering a TDD system, for intra-subnetwork communication, each channel will be divided into $K$ orthogonal OFDM subcarriers, where each subcarrier is occupied by one actuator and one sensor for downlink (DL) and uplink (UL) transmission so that there isn’t intra-subnetwork interference. For intra-subnetwork communication, the DL is from the controller to actuators and the UL is from sensors to controller. The controller continuously performs sensing the RSSI on every channel for every transmission time interval (TTI), and decides to select a suitable channel $\alpha_{i}(t)=\{1,2,...,M\}$ for intra-subnetwork data exchange. Here, RSSI is the total received power, which can be easily calculated in the communication system. The channel occupation indicator $\theta_{i}(m,t)$ is defined as
\begin{equation}
\label{eq6}
\theta_{i}(m,t)=\left\{
\begin{aligned}
1&,    &if~ \alpha_{i}(t)=m \\
0&,    &otherwise.  ,
\end{aligned}
\right.
\end{equation}
where $\theta_{i}(m,t)=1$ indicates the channel $m$ is selected by the subnetwork $i$ at time $t$.

When the subnetworks move, the interference is always changing. For intra-subnetwork transmission, every data packet is assumed to be mapped into a fixed payload, and the transmissions are periodically performed.
Per the reliability requirement of the subnetwork, the objective is to design the power control and resource management scheme in every TTI to maximize the subnetwork payload delivery probability within a time budget $T$ as
\begin{align}
P_r\left\{ \sum_{t=1}^{T}\sum_{m=1}^{M}\theta_{i}(m,t)\hat{C}_i(m,t) \geq \frac{B}{\Delta t} \right\} \, i \in \{1,2,...,N\},   
\end{align}
where ${\hat{C}_i(m,t)}$ is the channel capacity of the $m$-th channel at time $t$. \textit{B} denotes the size of payloads that are periodically generated by the local service of subnetworks in bits, and $\Delta t$ is the channel coherence time.

As the TDD frame structure is symmetric \cite{2020Distributed26}, we suppose that the DL channel capacity $C_i^{DL}(m,t)$ is the same as UL $C_i^{UL}(m,t)$, so that
\begin{align}
C_i^{DL}(m,t) = C_i^{UL}(m,t) = \hat{C}_i(m,t).
\end{align}

According to Shannon’s theorem, the maximal UL data rate $C_i^{UL}(m)$ between the controller and the $k$-th sensor is defined as
\begin{align}
C_i^{UL}(m,t) = W{\rm{log_2}}(1+\xi_{i}^{UL}(m,t)),
\end{align}
where $W$ is the bandwidth of each subcarrier for sensor/actuator. 
$\xi_{i}^{UL}(m)$ is the UL's Signal to Interference plus Noise Ratio (SINR) of the $i$-th subnetwork on channel $m$.

For the $i$-th subnetwork, the intra-subnetwork signal power received by AP on channel $m$ is computed as
\begin{align}
P_{i}^{UL}(m,t) = \sum_{k=1}^{K}P_{ik}^{se}(m)g_{ik}^{UL}(m)\theta_{i}(m,t),
\end{align}
where $P_{ik}^{se}(m)$ is the output power of the $k$-th sensor in the $i$-th subnetwork on channel $m$, and $g_{ik}^{UL}(m)$ is the channel gain between the controller and its $k$-th sensor at channel $m$. For each subnetwork, $P_{ik}^{se}(m)$ is a constant, and $g_{ik}^{UL}(m)$ can be calculated by channel estimation.

We use $RSSI_n$(m) to denote the RSSI of $m$-th subnetwork, and it consists of intra-subnetwork signal, inter-subnetwork interference and the noise. It can be calculated as
\begin{align}
RSSI_i(m) &= \sum_{k=1}^K P_{ik}^{se}(m)g_{ik}^{UL}(m)\theta_{i}(m,t) + \sum_{j=1, j \neq i}^N P_{j}^s(m,t)g_{ij}^{ss}(m)\theta_{i}(m,t) + \sigma^2.
\end{align}
Here $P_{j}^s(m,t)$ is the transmission power of $j$-th subnetwork on channel $m$. $g_{ij}^{ss}(m)$ denotes the instantaneous channel gain between the $i$-th and the $j$-th subnetworks, and $\sigma^2$ is the system random noise. It’s noteworthy that there is no direct communication between subnetworks, so the channel gain $g_{ij}^{ss}(m)$ is difficult to get, especially in the real industry environment.

Although the instantaneous inter-subnetwork channel gain ${g_{ij}^{ss}(m)}$ varies continuously due to the dynamic movement and can't be obtained in most subnetwork application scenarios, the UL SINR still $\xi_{i}^{UL}(m)$ can be calculated as
\begin{align}
\xi_{i}^{UL}(m) &= \nonumber
\frac{P_{ik}^{se}(m)g_{ik}^{UL}(m)}{\sum_{j=1,j \neq i}^N P_{j}^{s}(m)g_{ij}^{ss}(m)+ \sigma^2} \\&= \frac{\sum_{k=1}^K P_{ik}^{se}(m) g_{ik}^{UL}(m)\theta_{i}(m,t)}{RSSI_i(m)-{\sum_{k=1}^K P_{ik}^{se}(m) g_{ik}^{UL}(m)\theta_{i}(m,t)}} ,
\end{align}

After that, we can know the channel capacity. But for each subnetwork, without other subnetwork's information of transmission power, channel selection and channel gain, it’s really a challenging problem to learn the accurate action policy for collision avoidance only with its local RSSI.

\section{MARL Model for Resource Management of Subnetworks}\label{RL model}
In this work, we formulate the MARL problem as a partially observable Markov game (POMG) problem upon the wireless network, which extends the Markov decision process (MDP) to a multi-agent scenario. 
In this section, we first briefly introduce the POMG, and then formulate the mobile subnetwork environment as the multi-agent model, where each subnetwork acts as an agent. 

\subsection{POMG Model}

Here, we consider the POMG model in a wireless network environment, which extends the partially observable Markov decision process to a $N$-agent game. To avoid ambiguity, in the following, bold notations are used to denote global variables or joint variables. An $N$-agent Markov is formalized by a tuple $(N, S, \left\{ A_i \right\}_{i=1}^n, \left\{ R_i^t\right\}_{i=1}^n, T)$, in which $S$ denotes the state space, and $A_i$ indicates the action space of agent $i$, supposed to be the same for all subnetwork agents. The reward function of subnetwork agent $i$ is $R_i^t$. Let $A = A_1 \times A_2 \times \cdots \times A_n$ be the joint  action space. $R_i^t : S \times A \rightarrow R$ is the reward function of agent $i$. 
Besides, $T : S \times A \times S \rightarrow [0,1] $ is the state transition function, which determines the probability distribution of the next possible state, depending on the current states and actions of all agents. In addition, a partially observable problem is considered, where agent $i$ receives a local observation. 
Each agent learns a policy $\pi : O_i \rightarrow P(A_i)$ that maps local observations of subnetwork agent $i$ to a probability distribution of its actions. 
The ultimate objective of learning is to explore an optimal policy $\pi_i$, which can maximize their expected cumulative discounted rewards $J_i(\pi_i)$ that can be formulated as: 
\begin{align}
J_i(\pi_i) = \mathbb{E}_{a \sim \pi_1,...,\pi_N,s \sim T} \left[ \sum_{t=0}^{\infty} \gamma^t R_i^t(\boldsymbol{s_t},\boldsymbol{a_t})\right],
\end{align}
where $\gamma \in [0,1)$ represents the discount factor, and $s_t$ and $a_t$ represent states and actions of all subnetworks, respectively. Note that the optimal policy of the agent $i$, as well as the resulting $J_i$, 
are determined by the actions of all agents.

In a wireless network system, the non-stationarity of the network environment is greatly affected by the states and policies of neighbors, as the influence of each agent will indirectly spread to the whole network system. Therefore, to derive environmental stability, when selecting actions each mobile subnetwork must consider the influence of the behaviors of other subnetworks in a mobile subnetwork system.

\subsection{Environment Model}
Considering the scenario of resource management for subnetworks, as depicted in Fig. \ref{subnetworks}, multiple subnetworks compete to access limited spectrums with a certain power level, which can be modeled as a MARL problem. In the MARL model, the subnetwork takes an action in accordance with a policy and interacts with the communication environment. Firstly, the subnetwork agent observes a state $s^t$ belonging to the state space $S$. Then, an action $a^t$ (selecting power level and channel) is taken according to the policy $\pi$, leading to a new state  $s^{t+1}$ with a reward $r^t$. It is worth noting that our proposed MARL-based method is a centralized critic learning framework, but the learned policies are executed distributedly. 
The key elements of our MARL-based resource management scheme are described in detail as below.

\subsubsection{State and Observation Space}
In a real environment, the global states (e.g. global channel conditions  and the behaviors of all agents), are unknown for individual subnetwork agents. 
With the observation function, each subnetwork agent can only learn the knowledge of the physical environment through the limited observation space. 
It is difficult for a subnetwork to know other agents' transmission power and channel gain, but easy to obtain the overall interference power at each spectrum band. 
Therefore, in our approach, the observation space of a subnetwork agent $i$ contains the RSSI at per channel, where RSSI is composed of the known traffic signal and the total interference from all other agents.
Compared with other existing MARL-based algorithms, in our approach, RSSI only offers the partial inference information. 
Meanwhile, the remaining traffic payload $B_i$, the action $A_i^{(t-1)}$ at the previous time step, and the remaining time budget $T_i$ are contained in the agent's local observation space to better capture the queuing status of each subnetwork. Therefore, the observations $O_i(S_t)$ of an agent $i$ at time $t$ can be summarized as:
\begin{align}
O_i(S_t) = \{A_i^{(t-1)}, B_i, T_i, \{RSSI_i(m)\}_{m\in \{1,2,...,M\}}  \}.
\end{align}

\subsubsection{Action}
The resource management design of subnetworks is essentially the channel selection and power control of subnetworks. In other words, the subnetwork takes an action, which is composed of channel selection and power control, based on the policy trained by our approach. In this work, subnetwork transmission power takes a discrete value, which is limited to three levels, namely, [10, 0, -114] dBm. 
Note that there will be no transmission power when -114 dBm is chosen. Consequently, the action space dimension is computed as $M \times \beta$ when the number of sets of optional channels is $M$ and there are $\beta$ levels transmission power. 

\subsubsection{Reward Function Design}

An appropriate reward function should be designed in line with the specific task to ensure the task to be well accomplished in the MARL problem. Each agent is expected to make decisions that could maximize rewards based on the interaction with the environment. Therefore, we design a tailored reward function to solve the subnetwork resource management problem as described in Section \ref{system model}. Our goal is to maximize the chance of successfully accomplishing payload transmission within a certain time constraint $T$. 
To achieve this goal, for each subnetwork agent $i$, if there are still remaining payloads $B_i$ to be transmitted at time $t$, the reward $R_i^t$ is calculated as the sum of transmission rate $\hat{C}_i^t$ of channels selected by the subnetwork divided by the target payload $L_i$. Otherwise, the reward will be a constant value $\eta$, which is greater than the maximum possible subnetwork transmission rate. Therefore, the reward function $R_i^t$ is expressed as:
\begin{equation}
\label{eq6}
R_i^t=\left\{
\begin{aligned}
\sum_{m=1}^M \theta_{i}(m,t)\hat{C}_i^t(m)/L_i & , & if~ B_i \geq  0, \\
\eta&,   &otherwise.
\end{aligned}
\right.
\end{equation}
In this MARL model, the objective of learning is to obtain an optimal policy $\pi_{*}$, which can map the states in state space to the probabilities of actions in action space, targeting at maximizing the expected cumulative reward from an initial state $s$. A larger cumulative reward translates into a faster transmission rate and a smaller link outage ratio.

\section{MARL based on GNN and Attention for Resource Management} \label{approach}
In this paper, we aim to solve the cooperative resource management problem of a subnetwork system, which can be formulated as a MARL problem. To overcome the intrinsic non-stationarity of multi-agent environment, inspired by the prior works about deep MARL \cite{2017Multiagent29}\cite{2017Counterfactual30}\cite{2018Value31}\cite{2018QMIX32}, 
a centralized training and distributed execution paradigm is adopted in our solution. 
Our approach is an extension of SAC \cite{2018Soft34}, where each subnetwork is treated as an autonomous agent with two networks, critic and policy, just like the multi-agent algorithm MADDPG \cite{2017Multi33}. 
Our approach allows subnetworks to utilize other subnetworks' states and actions to train policies, which is conducive to overcoming the instability of the multi-agent environment and coordinating the cooperation among subnetworks, during the centralized training process. To leverage the powerful coding capability of GNN, we first formulate the interference relationships between subnetworks as a graph model $G$, in which each vertex denotes a subnetwork agent and the potential interference relationship between mobile subnetworks are indicated as edges. Intuitively, different subnetwork agents have distinct importance to other agents. For example, if a subnetwork node $i$ has selected the same channel as another node $j$, then the node $j$ obviously has higher importance than other nodes. 
To reason the relations between all subnetworks, which are used to simplify the learning process, in this work, a new attention-based GNN, named GA-Net, is developed. GA-Net can be leveraged to decouple and reason potential relationship among subnetworks from the historical information of mobile subnetworks, which will contribute to learning a critic for each subnetwork agent through selective attention to features of other subnetwork agents. 

In addition, to avoid excessive computational overhead during the training process, the centralized training process is shifted to the BS, where each agent's critic, as an evaluator of action quality, can be augmented with additional historical information of other subnetworks uploaded to BS. The historical information of subnetworks provides a richer basis for reasoning various potential relationships among subnetworks from RSSI.
After the completion of centralized training, a subnetwork $i$ fetches the trained parameters of the target policy network from BS, and integrates them into its policy network $i$ in the distributed execution phase. Then, the policy $i$ chooses the optimal action (selection of channel and power level) $a_i^t$ according to the state $s_i^t$, which is obtained by the subnetwork $i$. The environment will return a reward $R_i^t$ when the subnetwork $i$ takes the action $a_i^t$. 

\begin{figure}[htb]  
	\centering  
	\includegraphics[width=13cm]{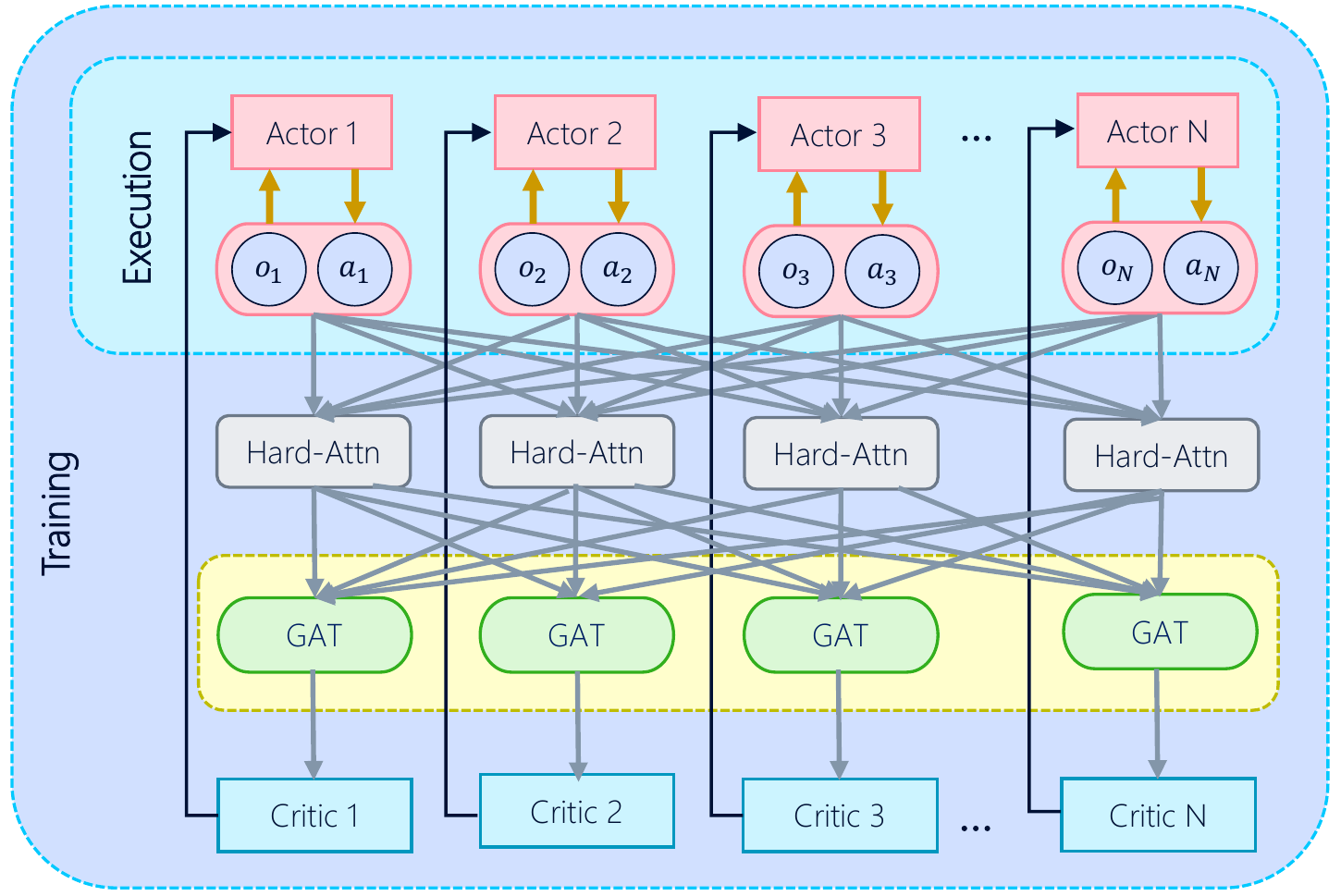}  
	\caption{An overview of our proposed framework.} \label{overview}
\end{figure}

Furthermore, based on the attention network and GNN, we put forward a novel game scheme, named GA-Net.
In practical terms, GA-Net formalizes the game states as a graph and computes the node state embedding of each subnetwork to represent the subnetwork’s status by decoupling and reasoning various potential relations among subnetworks from RSSI. In addition, the historical information of subnetworks is beneficial to fully evaluate the potential impact among subnetworks.
Specifically, GA-Net is composed of an improved hard attention layer and a graph attention network (GAT)  \cite{velivckovic2017graph} layer enhanced with multi-head attention, where the improved hard attention is utilized to exclude the impact of unrelated subnetworks to decrease the computational complexity during training, and the multi-head attention is leveraged to decouple and reason various potential relationships among subnetworks to obtain an information-condensed joint state encoding using GNN. 
Moreover, our approach extends the SAC framework to multi-agent environments and employs GA-Net to train a critic for each agent, aiming to learn how to cooperate to optimize resource management. Fig. \ref{overview} illustrates the overview of our approach.

\subsection{GA-Net Based Attention} \label{GA-Net}
Inspired by the attention mechanism, we decouple and reason various potential relationships among subnetworks through paying selective attention to various features of other subnetworks, which is conducive to training an augmented critic for each subnetwork.

In this context, we propose the GA-Net game algorithm, which is beneficial to decrease the computing complexity and reason the potential relationships among subnetworks by the aid of an improved hard-attention and multi-head self attention. 
The potential inference relationships among subnetworks are modeled as a graph model $G$. We consider a partially observable scene, where each agent $i$ receives a current observation $o_i^t$ at each time step $t$. First, the current observation $o_i^t$ of subnetwork $i$ is encoded into state encoding $s_i^t$ by multilayer perceptron (MLP).  The current state code $s_i^t$ of the subnetwork $i$ and its state code at the previous K times are encoded to $e_i^t$ by Gate Recurrent Unit (GRU). Then, the state encoding $e_i^t$ is used to learn the interference relationships among subnetworks by GA-Net networks, resulting in a high-level state encoding, which fuses contributions from other subnetworks. To simplify the representation, $o_i^t$, $s_i^t$ and $e_i^t$ are simplified as $o_i$, $s_i$ and $e_i$, respectively. 
Notably, $e_i$ is fed into GAT and hard attention as input. On the one hand, as hard attention can generate a one-hot vector as an output, thus we can train a hard attention layer to get which subnetworks have potential interference relationships with each other. 
Through the hard attention mechanism, the relationships among subnetworks are simplified and a sub-graph $G_i$ for subnetwork $i$, in which only the subnetworks needing to interact with subnetworks $i$ are connected within the graph $G_i$, can be obtained. On the other hand, each subnetwork has a different degree of relevance to a specific subnetwork, which means each edge of the graph $G_i$ has different weights. 
At the same time, the interference between subnetworks can be affected by multiple factors and multi-head attention can be utilized to extract various representation from various state feature sub-spaces. Therefore, for sub-graph $G_i$, a GAT layer with multi-head attention is trained to learn the weights of subnetworks to subnetwork $i$ in different state feature sub-spaces, obtaining the joint state encoding of agent $i$ that contains the contributions of all other subnetworks to the subnetwork $i$. Through joint state encoding, better decisions can be achieved in our approach. 

\begin{figure*} \centering
	\subfigure[The network structure of hard attention.] { \label{hard_attention}
		\includegraphics[width=0.37\linewidth]{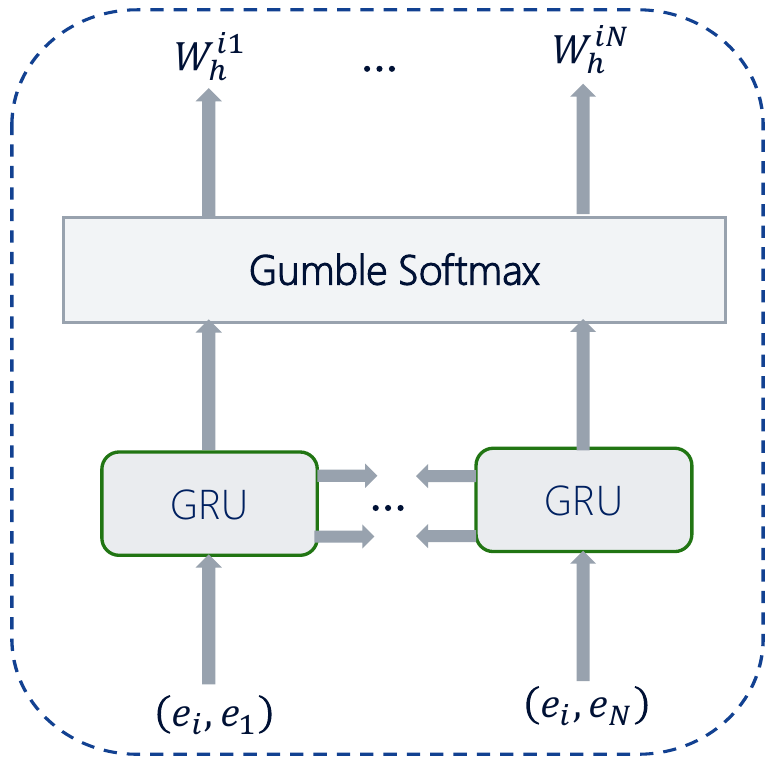}
	}
	\hspace{0.5cm}
	\subfigure[The network structure of multi-head attention.] { \label{multi-head_attention}
		\includegraphics[width=0.5\linewidth]{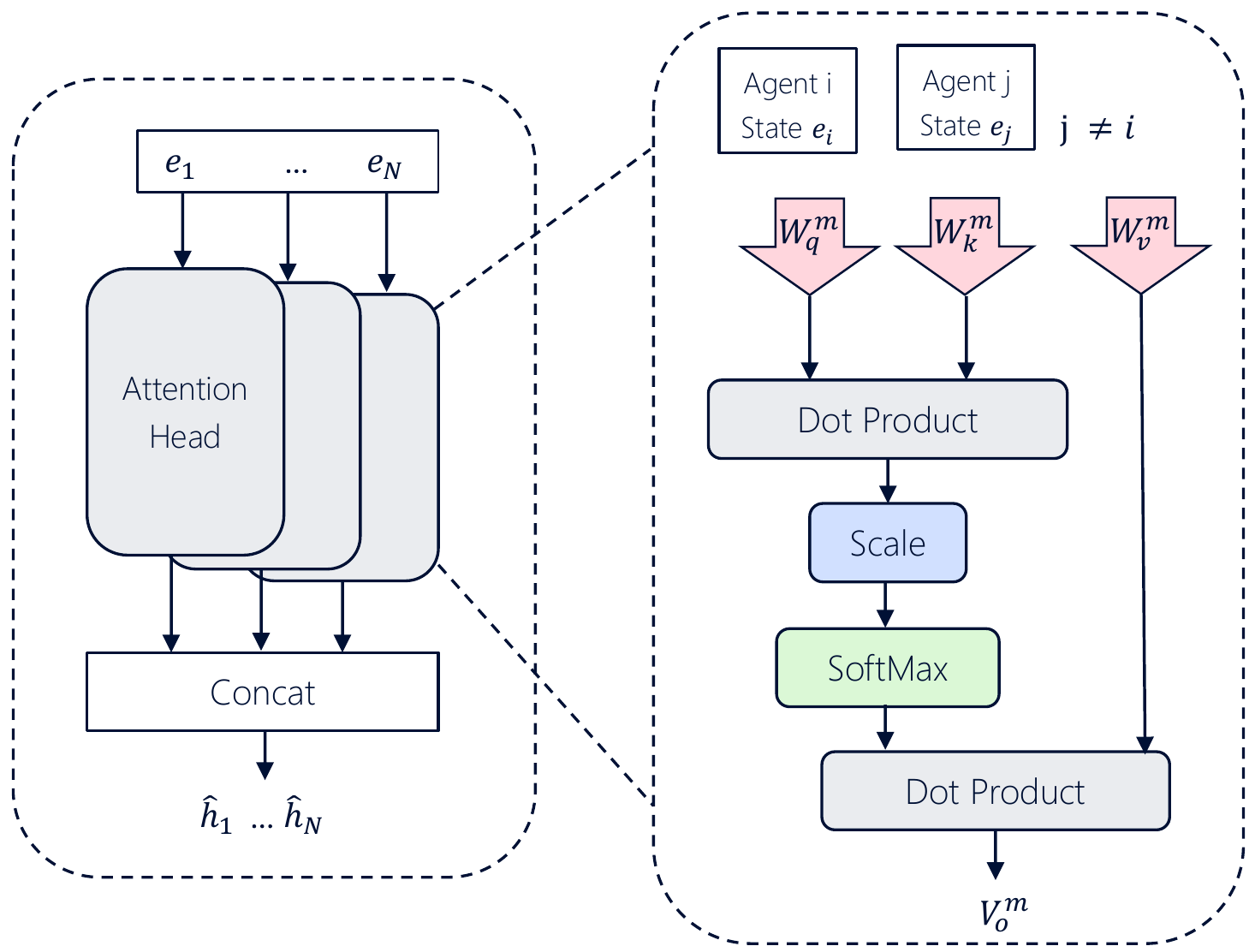}	}
	\caption{ The network structure of hard attention and multi-head attention}
	\label{fig}
\end{figure*}

In this work, the hard-attention, as depicted in Fig. \ref{hard_attention}, uses a GRU to achieve the weights of edges, which determine whether there is an interaction between subnetworks, and outputs a weight $W_h\in \{0,1\}$ for the edge of the subnetwork in each time step. 
Specifically, for subnetwork $i$, we first input the state encoding $e_i$ and $e_j$ fused with historical information into the GRU network layer following 
\begin{align}
h_{ij} = h(GRU(e_i, e_j)), & 
\end{align}
where $h(\cdot)$ is a fully connected layer for state encoding.
Traditional GRU networks cannot make full use of all subnetworks' feature information due to a reasonable and short-sighted fact that the sequence of subnetworks plays a pivotal role in the procedure. Therefore, we use the bidirectional GRU (BiGRU) model, instead of traditional GRU, so that the relationship between subnetwork $i$ and $j$ also depends on states of other subnetworks. In addition, as the hard-attention often cannot achieve back-propagation of gradients, we use gumbel-softmax function to address it. Hence, the hard-attention weight value $W_h^{i,j}$ of the edge linking agent $i$ and $j$ can be calculated as 
\begin{align}
W_h^{i,j} = gumbel(h_{ij}), & 
\end{align}
where $gumbel(\cdot)$ indicates the gumbel-softmax function. 
For each subnetwork $i$, the hard-attention mechanism can be leveraged to help derive a sub-graph $G_i$, in which subnetwork $i$ only connects with the subnetworks that can help the subnetwork $i$ choose the optimal action. 

Subsequently, for subnetwork $i$, the GAT with multi-head attention is employed to derive a joint state encoding. First, multi-head attention, as shown in Fig. \ref{multi-head_attention}, is used to compute the weights of other subnetworks for subnetwork $i$. Specifically, the $m$-$th$ head uses a separate set of parameters ($W_k^m, W_q^m, W_v^m$) to compute a `query' $Q_i^m = W_q^m e_i$, a `key' $K_i^m = W_k^m e_i$ and a `value' $V_i^m = W_v^m e_i$. After a query-key pair ($Q_j^m, V_j^m$) from another subnetwork $j$ in $G_i$ is received, the subnetwork $i$ aggregates the states encoding of other subnetworks with the weight $w_{i,j}^{m} = softmax(\frac{Q_i^m(K_i^m)}{d_k})$. The $d_k$ represents the dimensionality of the key $K_i^m$, which is used to prevent the gradient from vanishing. Note that the parameters ($W_k^m$, $W_q^m$, $W_v^m$) are shared across all subnetworks. Then, we can use GNN to compute an aggregated contribution encoding $V_o^m$  from all other subnetworks in $G_i$ through calculating a weighted sum of the `values' of other subnetworks. Ultimately, the contributions encoding from all heads are concatenated as a state embedding $\hat{h_i}$:
\begin{align}
\hat{h_i} = ||_{m=1}^L \sigma(\sum_{j \in G_i} w_{ij}^mW^ms_i) \label{other_contribution},
\end{align}
where $W^m$ denotes trainable parameters, $L$ represents the number of attention heads, $\sigma$ is a nonlinear function, and $||$ represents the connecting state encoding.

To sum up, through the hard attention layer, we get a simplified graph, where each subnetwork only connects to the subnetworks that need to interact with certain importance weight in the graph $G$, which excludes the impact of unrelated subnetworks and is conducive to decreasing the computing complexity. For each subnetwork $i$, we utilize GAT with multi-head attention to achieve an aggregated contribution encoding from all other subnetworks through calculating a weighted sum of the values of other subnetwork in $G_i$ and concatenate the contributions encoding from all heads as a single vector $\hat{h_i}$. Eventually, the subnetwork $i$ updates its state encoding by using a neural network to conduct a non-linear transformation of the current state encoding cascaded with $\hat{h_i}$.

\subsection{Learning Critics With GA-Net}
In order to compute the Q-value function $Q_{\theta_i}(\mathbf{s},\mathbf{a})$ for subnetwork agent $i$, 
the observations $\mathbf{s}=(s_1,...s_N)$ and actions $\mathbf{a}=(a_1,...a_N)$  are used as input to the critic for all subnetworks. $Q_{\theta_i}(\mathbf{s},\mathbf{a})$ is a value function of subnetwork agent state that combines observation and action with other subnetwork agents' contribution $\hat{h_i}$:
\begin{align}
Q_{\theta_i}(\mathbf{s},\mathbf{a}) = h_i(f_i(s_i, a_i), \hat{h_i}), 
\end{align}
where $h_i$ and $f_i$ are MLPs. $\hat{h_i}$ is calculated using Eq. (\ref{other_contribution}), which is state encoding based on GA-Net.

Through minimizing the joint regression loss function, the critics of all subnetwork agents are updated together following
\begin{align}
\mathcal{L}(\theta_i) = \sum_{i=1}^N   \mathbb{E}_{(\mathbf{s},\mathbf{a},\mathbf{r},\mathbf{s^{\prime}})\sim{D}}[(Q_{\theta_i}(\mathbf{s},\mathbf{a})-y_i)^2],
\end{align}
where
\begin{align}
y_i  = r_i + \gamma \mathbb{E}_{\mathbf{a^{\prime}} \sim {\mathbf{\pi_{\bar{\psi}}}(\mathbf{s^{\prime}})}}[Q_{\bar{\theta}_i}(\mathbf{s^{\prime}}, \mathbf{a^{\prime}})-\alpha log(\pi_{\bar{\psi_i}}(a_i^{\prime}\lvert s_i^{\prime}))],
\end{align}
where $Q_{\bar{\theta_i}}$ and $\pi_{\bar{\psi_i}}$ represent the target critic and target policy, respectively. $\alpha$ is the temperature parameter, which is used to balance the maximum entropy and reward. The individual policies are updated by gradient ascent: 
\begin{align}\label{policy_loss}
\nabla_{\psi_i}J(\pi_{\psi_i}) = 
&\mathbb{E}_{\mathbf{s} \sim D, \mathbf{a} \sim \pi}[\nabla_{\psi_i}log(\pi_{\psi_i}(a_i \lvert o_i))(-\alpha log(\pi_{\psi_i}(a_i \lvert s_i))) +Q_{\theta_i}(\mathbf{s},\mathbf{a}) - b(\mathbf{s},\mathbf{a_{\backslash i}})],
\end{align}
where $b(\mathbf{s},\mathbf{a_{\backslash i}})$ indicates the baseline that is used to compute the advantage function. 
It is noteworthy that aiming to compute the gradient estimation of subnetwork agent $i$, all actions $\mathbf{a}$ are sampled from the current policies of all subnetwork agents rather than the replay buffer, and the gradient estimation can be coordinated according to their current policies.

To address the multi-agent credit assignment problem, we compute an advantage function that utilizes the baseline by marginalizing out the specific action of the agent from $Q$-value.
That is, when the actions of all other agents are fixed, by comparing the value of a specific action with the average value of all possible actions of the agent, we can know whether the specific action will result in an increase in the expected reward, or whether the increase in reward is attributable to the actions of other agents.
The advantage function is defined as below:
\begin{align}
A_i(\mathbf{s},\mathbf{a}) = Q_{\theta_i}(\mathbf{s},\mathbf{a}) - b(\mathbf{s},\mathbf{a_{\backslash i}}),
\end{align} 
where
\begin{align}
b(\mathbf{s},\mathbf{a_{\backslash i}}) = \mathbb{E}_{a_i \sim \pi_{\psi_i}(o_i)}[Q_{\theta_i}(\mathbf{s}, (a_i, a_{\backslash i}))].
\end{align}

\subsection{Training and Execution}
In this work, the policy is trained using the Soft Actor-Critic algorithm, which is an off-policy and actor-critic method. Owing to that the mobile subnetwork has weaker computing power than the BS, the training process of our algorithm is centrally conducted at the BS. 
Thus, only the trained parameters of the target policy network need to be downloaded from the base station, and the subnetwork only follows the policy to carry out during the phase of distributed execution. The training algorithm is shown in Algorithm \ref{algorithm}. 

\begin{algorithm}[h]
	\caption{Training Algorithm} \label{algorithm}
	\LinesNumbered
	\KwIn{Initial parameters of policy and critic network.}
    \KwOut{The parameters of target policy network.}
    Initial replay buffer, $D$;\\
    \For{episode $\in [1,..n]$}{
      Reset the environment with the initial state, and get initial $s_i$ for each agent $i$\;
      \For{$t \in [1...T]$ steps per episode}{
        Sample actions $a_i^t \sim \pi(\cdot \lvert s_i^t)$ for each subnetwork, $i$\;
        Execute actions in environment and get $s_i$ $r_i$ for all subnetworks\;
        Store the tuples $(\textbf{s}^t,\textbf{a}^t,\textbf{r}^t,\textbf{s}^{t+1})$ for all environments in D\;
        Sample a random minibatch B from replay buffer D\;
        Update the critic network by minimizing the loss using $\nabla$$ L_Q(\theta)$ and Adam \cite{2014Adam22}\;
        Update the policy network applying the sampled policy gradient and Adam based on Eq. (\ref{policy_loss})\;
        Update target parameters: \\
        $\overline{\theta} = \tau\overline{\theta} + (l-\tau)\theta$, 
        $\overline{\psi} = \tau\overline{\psi} + (l-\tau)\psi$.
      }
    }
\end{algorithm}

\begin{algorithm}[h]
\caption{Execution Algorithm}
\label{execution}
\LinesNumbered 
\KwIn{        
    Target policy network parameters $\overline{\psi}$;}
The actor networks of all subnetworks load network parameters from target actor netwroks;

All subnetworks receive their initial observation state $\boldsymbol{s_0}=\{s_0^1,s_0^2, s_0^3 ...\}$;

\For{$t \in [1...T]$ steps}
{
All subnetworks select actions
$\boldsymbol{a^t} = \{a_0^t, a_1^t, a_2^t...\}$ according to the actor networks\;
All subnetworks execute $\boldsymbol{a^t}$ and get the current reward $\boldsymbol{r^t}$ and the new state $\boldsymbol{s^t}$.
}
\end{algorithm}

The execution algorithm is shown in Algorithm \ref{execution}. The subnetworks first download the parameters of target networks from the BS and load parameters of the target networks to the actor networks. Then, the actor networks of all subnetworks select actions with observation as input.

\section{Performance Evaluation}\label{evaluation}
To validate the effectiveness and efficiency of our proposed MARL-based resource management approach, in this section we conduct comprehensive experiments and compare with both traditional approaches and DRL-based approaches.


\subsection{Simulation Settings}
\begin{figure}[htb]  
	\centering  
	\includegraphics[width=9cm]{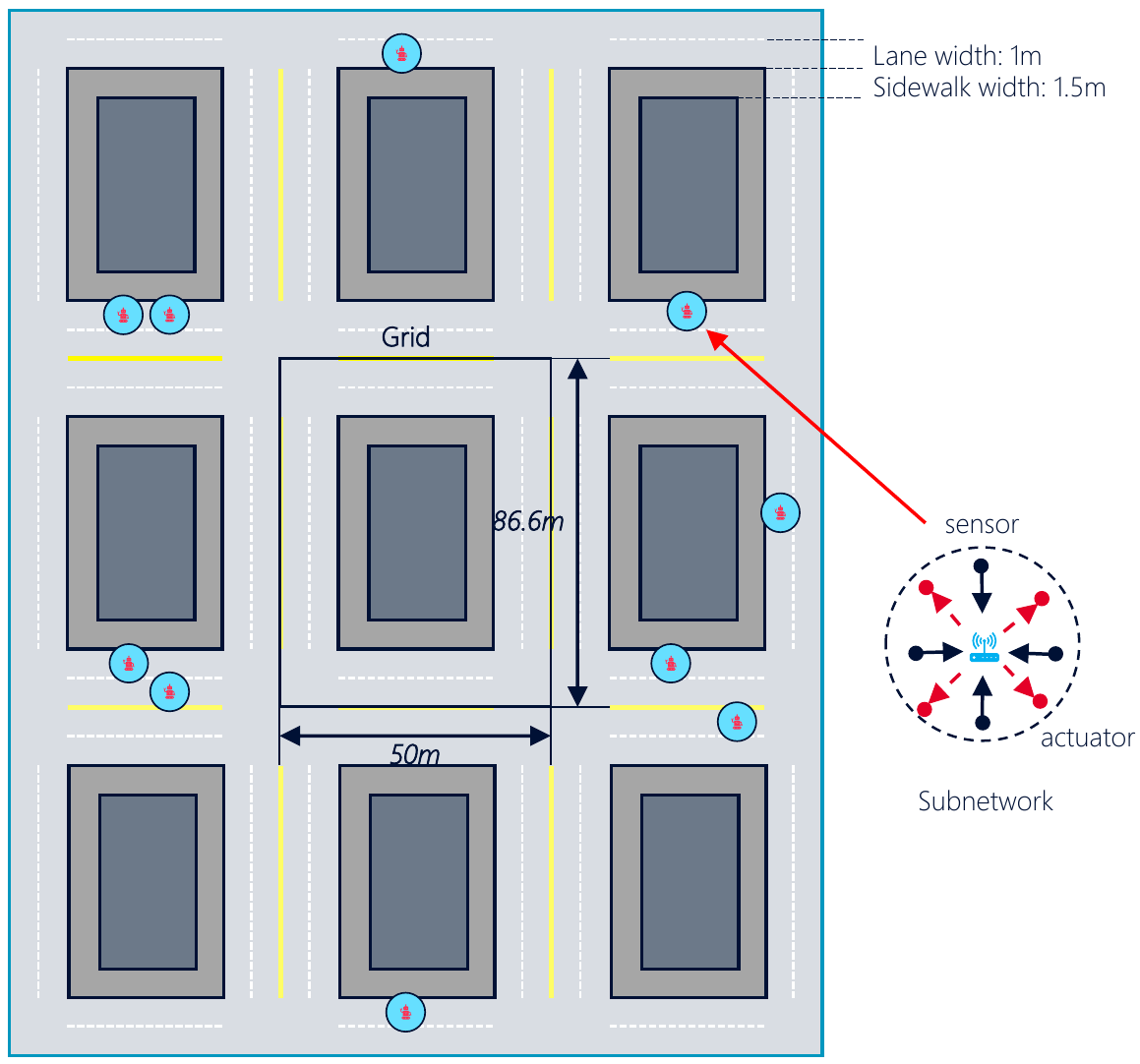}  
	\caption{The simulation environment.} \label{simulation}
\end{figure}

We consider a network with multiple subnetworks in our simulation environment, as shown in Fig. \ref{simulation}, which is similar to the case defined in 3GPP-36.885. The available channels are assumed to be limited. For each transmission of subnetworks, a fixed payload is mapped into a single OFDM channel. The simulation parameters are summarized in Table \ref{notations}.

We use the indoor to indoor channel model defined in 3GPP-38.901 for simulation, where fast fading varies with time steps. Each subnetwork moves along the road at a random speed $v$ = 2 $\sim$ 3 m/s. 
When the subnetworks reach the crossroad, there is 50\% chance to go straight, and 25\% chance to turn left or right, respectively.
The TTI is set to be 1 ms.

The state encoder of the subnetwork takes 32-dim states as output. Both actor and critic networks in our framework are three ReLU fully connected layers, where two hidden layers have 64 and 32 neurons, respectively. The attention module uses 32-dimensional queries, keys, and values to derive attentions. All parameters are trained using Adam Optimizer.  
The learning rates of actor and critic are set to be 0.0001 and 0.001, respectively.  
The discount factor of reward is set to be 0.9.  The receiver's noise figure is 5 dB. The resource allocation policy is trained using the Algorithm \ref{algorithm} on a virtual machine with an Nvidia Tesla P100 graphics card on the Google Colab platform \cite{googlecolab25}.

\subsection{Compared Algorithms}
To fully validate the effectiveness of the proposed approach, we compare our approach with other four approaches: i) \textit{Random Baseline}, which selects the channel and transmission power for each subnetwork in a random manner at each time step; ii) \textit{Dynamic Greedy Algorithm (DGA)} \cite{2020Distributed26}, which selects the channel with least RSSI every time interval, and the transmission power is always the maximal. iii) \textit{AC}\cite{2015Continuous}, which is the most classic deep RL method and can handle continuous action space; iv) \textit{MADDPG} \cite{2017Multi33}, which can obtain global knowledge and deal with discrete action spaces with Gumbel-Softmax;  v) \textit{GA-Net w/o Hard Attn}, which is GA-Net without hard attention; vi) \textit{GA-Net w/o Attn}, which is GA-Net without hard attention and soft attention; vii) \textit{QMIX}, which is a state-of-the-art multi-agent reinforcement learning algorithm based on value function decomposition.

\subsection{Experimental Results}
The evaluation results will be demonstrated with respect to various performance metrics, including the convergence speed, episode reward, outage probability of subnetwork. In addition, we provide an in-depth analysis of the channel and power level selection directed by the policy learned through our approach.
\begin{itemize}
	\item Convergence speed: the rate of convergence to the minimum return during training.
	\item Episode reward: the reward of each episode.
	\item Outage probability of subnetwork: the probability of data transmission failure. It is evaluated under two different scenarios: variable subnetwork densities with fixed channel bandwidth, and fixed subnetwork density with variable channel bandwidth.
	\item Channel and power selection: this reveals what GA-Net based MARL learns from the RSSI to a specific action.	
\end{itemize}

\begin{table}[]	
	\centering
	\caption{Summary of Notations}\label{notations}
	\begin{tabular}{|l|l|}
		\hline
		Paramter & Value \\ \hline
		Deployment area [$m^2$] & $259.8 \times 150$   \\ \hline
		Minimum inter-subnetwork distance [m]  &  1.5           \\ \hline
		Velocity, $v$[m/s] &  2.0 $\sim$ 3.0 \\ \hline
		Probability of going straight  & 50$\%$ \\ \hline
		Probability of turning left  & 25$\%$ \\ \hline
		Probability of turning right & 25$\%$ \\ \hline
		Transmission Time Interval [ms] & 1.0 \\ \hline   
		Snapshot duration [ms] & 100 \\ \hline
		Noise power $\sigma^2$ [dBm] & -114 \\ \hline
		Subnetwork transmission power [dBm] & [10, 0, -114] \\ \hline
		Tx antenna gain [dBi] & 4 \\ \hline      
		Rx antenna gain [dBi] & 4 \\ \hline
		Rx noise figure [dB] & 5 \\ \hline
		Channel model & Indoor to Indoor \\ \hline
	\end{tabular}
\end{table}

\begin{figure}[htb]  
	\centering  
	\includegraphics[width=9cm]{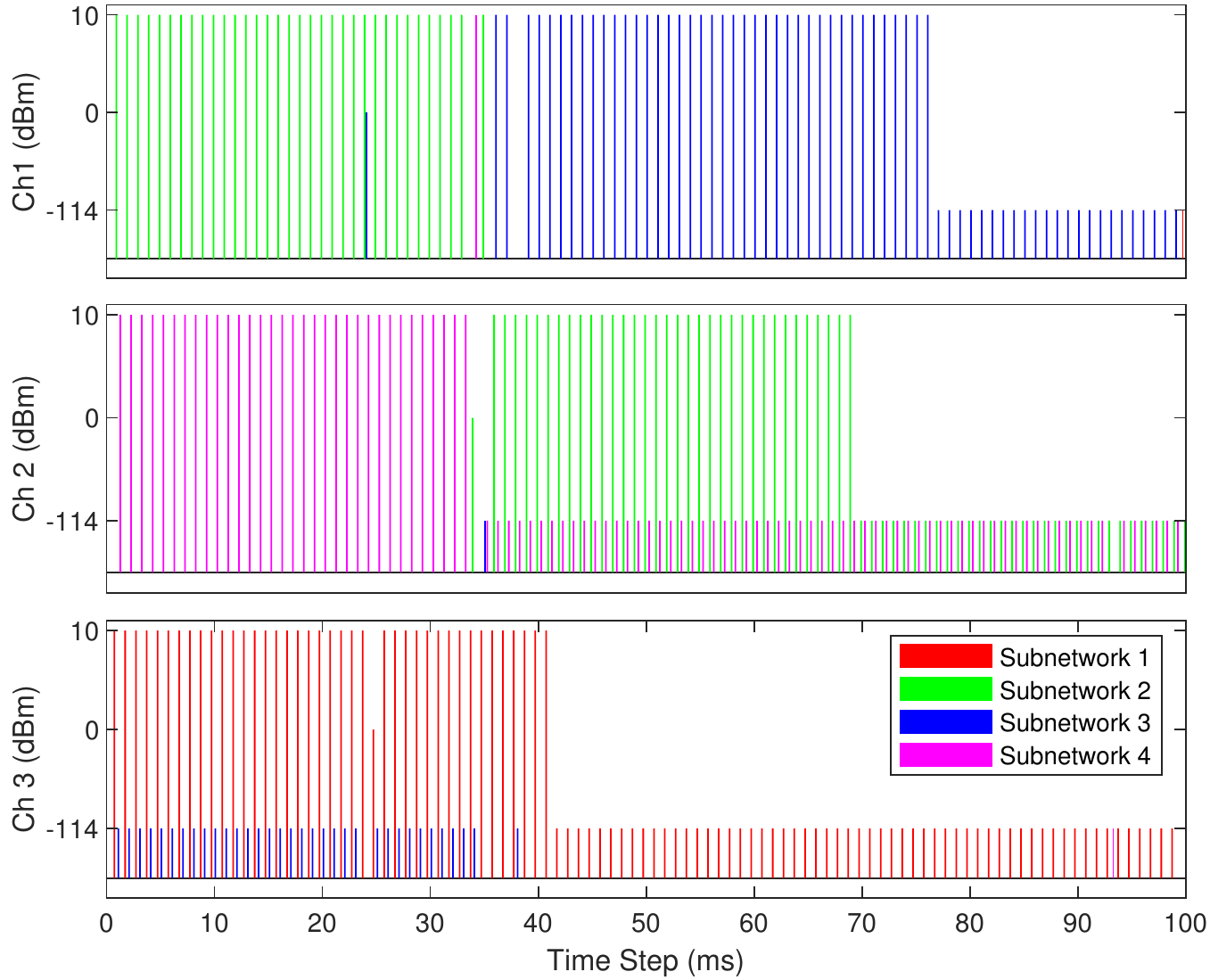}  
	\caption{Channel and power selection within coherence time.} \label{action_power}
\end{figure}
\begin{figure}[htb]  
	\centering  
	\includegraphics[width=8.5cm]{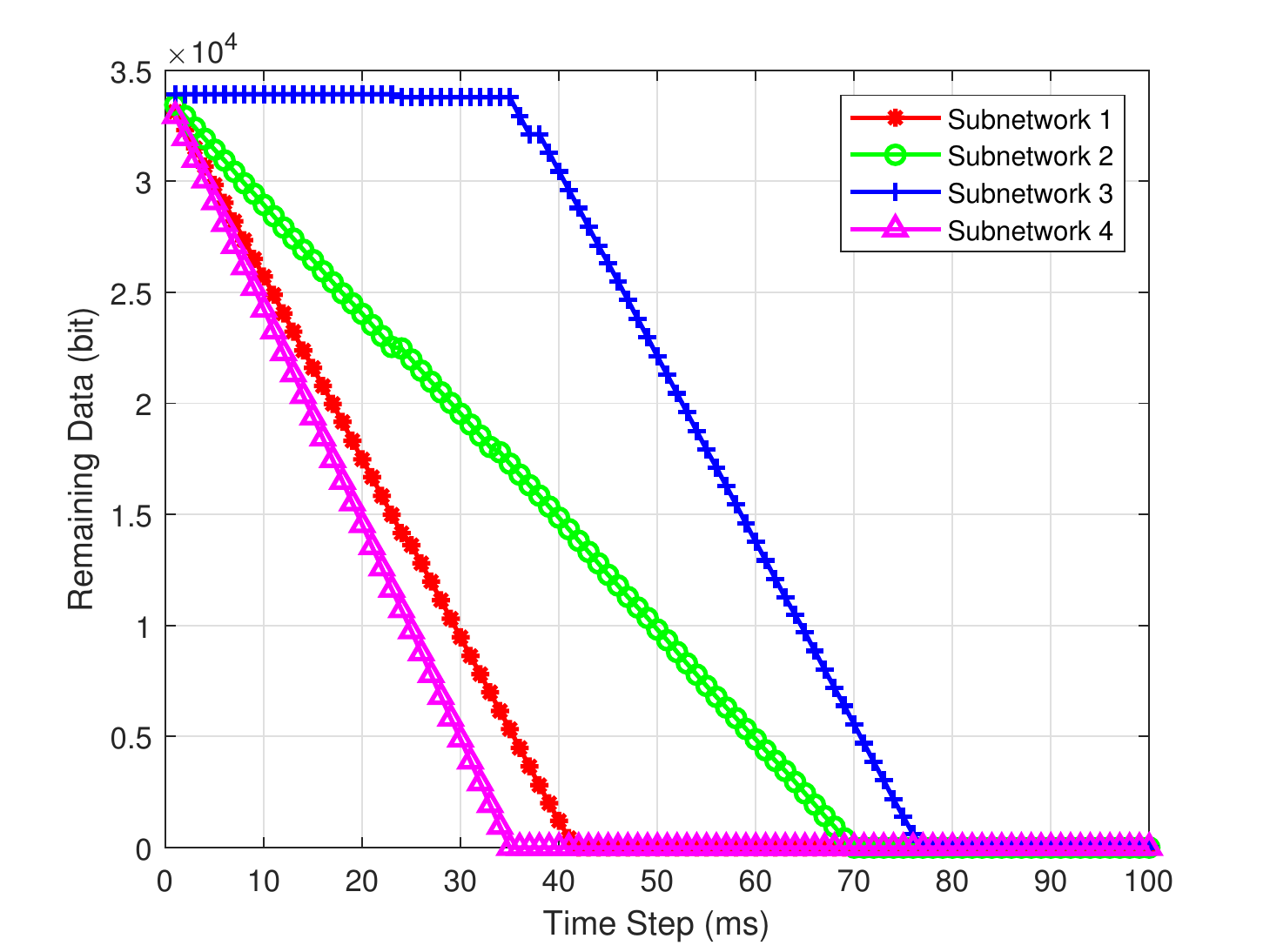}
	\caption{The remaining payload of each subnetwork within an episode.} \label{remaining_data}
\end{figure}

First, we will analyze the coordination process of policy trained by our approach, which is used to select channel and power level for multiple subnetworks within coherence time.
To analyze how multiple subnetworks coordinate the selection of channel and power level to maximize resource utilization, we select an episode, in which the policy trained by our approach enables all subnetworks to successfully deliver the data. Moreover, for the sake of clearer demonstration, we select a scenario with 4 subnetworks and 3 channels, the target payloads are all 34K bits.
Fig. \ref{action_power} shows the experimental results of channel and power selection within coherence time, i.e, $T$=100ms, for all subnetworks. It can be observed that subnetwork $2$ and $3$ select one channel separately, which is not shared with other subnetworks in the first $35$ time steps, and select the maximum power level. 
Meanwhile, subnetworks $1$ and $4$ choose the same channel as the number of channels are less than the number of subnetworks. However, subnetwork $1$ selects the maximum power level while subnetwork $4$ selects the minimum power level, which leads to the least interference and makes better usage of the channel to get the maximum total throughput. At $35$-th time step, subnetwork $4$ adjusts its own power level to the minimum, encouraging other subnetworks with the same channel to transmit data with higher data rate. While subnetwork $3$ preempts channel $1$, subnetwork $2$ occupies the channel $2$.
Fig.\ref{remaining_data} depicts the detailed remaining payload within an episode as shown in Fig. \ref{action_power}.
It can be seen that the payload transmission of subnetwork 4 is finished earlier at the $35$-th time step. Then, subnetwork 1 and 2 complete data transmission successively. Ultimately, subnetwork 3 is the latest one to complete payload transmission. From another perspective, subnetwork 1, 2, and 4 remain in data transmission until the payload drops to 0, however, subnetwork 3 does not start data transmission in the early time step until subnetwork 4 completes transmission. In conjunction with the observations revealed in Fig. \ref{action_power} and  Fig. \ref{remaining_data}, we can find that all subnetworks have learned a more beneficial policy by leveraging our approach and RSSI such that channels can be fully utilized and the interference is reduced as much as possible.

\begin{figure}[htb]  
	\centering  
	\includegraphics[width=8.5cm]{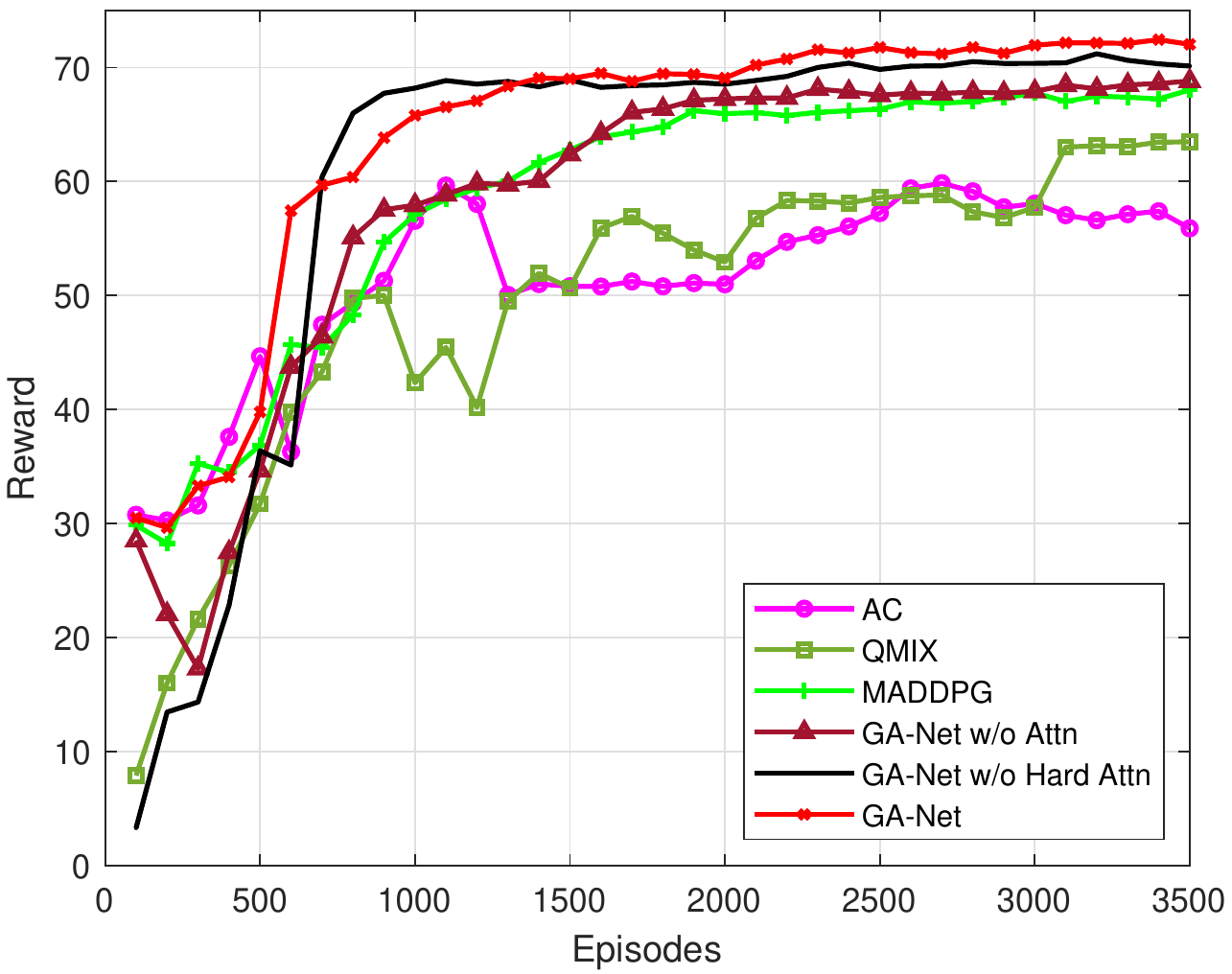}  
	\caption{Comparisons of the cumulative rewards per training episode.} \label{reward}
\end{figure}

Furthermore, we also evaluate the convergence performance of various algorithms. 
Fig. \ref{reward} compares the cumulative rewards per training episode during the training process to investigate the convergence performance of the compared three RL-based method with 8 subnetworks and 6 channels. 
Intuitively, a good reward function results in an efficient resource management policy, which allows subnetworks to cooperate for faster data transmission.
In the training phase, the AC based method has the worst performance on the cumulative reward and convergence speed. 
This is mainly due to the fact that each agent concentrates more on its own environment and reward, without considering the impact of the global environment on the stability of the training process and the impact of cooperation among multiple agents on system performance. 
Moreover, it reveals that MADDPG achieves faster cumulative and better stability than AC, as MADDPG uses the state and action information of all subnetworks to assist in training for obtaining a cooperative policy for all subnetworks. In addition, as revealed in Fig. \ref{reward}, our proposed algorithm (GA-Net) converges the fastest to the stable reward outperforming  AC, MADDPG, QMIX, GA-Net w/o Hard Attn and GA-Net w/o Attn, and the Convergence reward is greater than other algorithms. 
Because the state input is only RSSI instead of detailed channel gain, each agent cannot get the full interference information to take exact actions, thus it should analyze the importance of the interference under the partial observation. 

The better performance can be attributed to the fact that the resource management policy trained by our GA-Net approach can extract more critical features, with fewer noise, from other subnetworks that are possible to have stronger latent interference. 


\begin{figure}[htb]  
	\centering  
	\includegraphics[width=8.5cm]{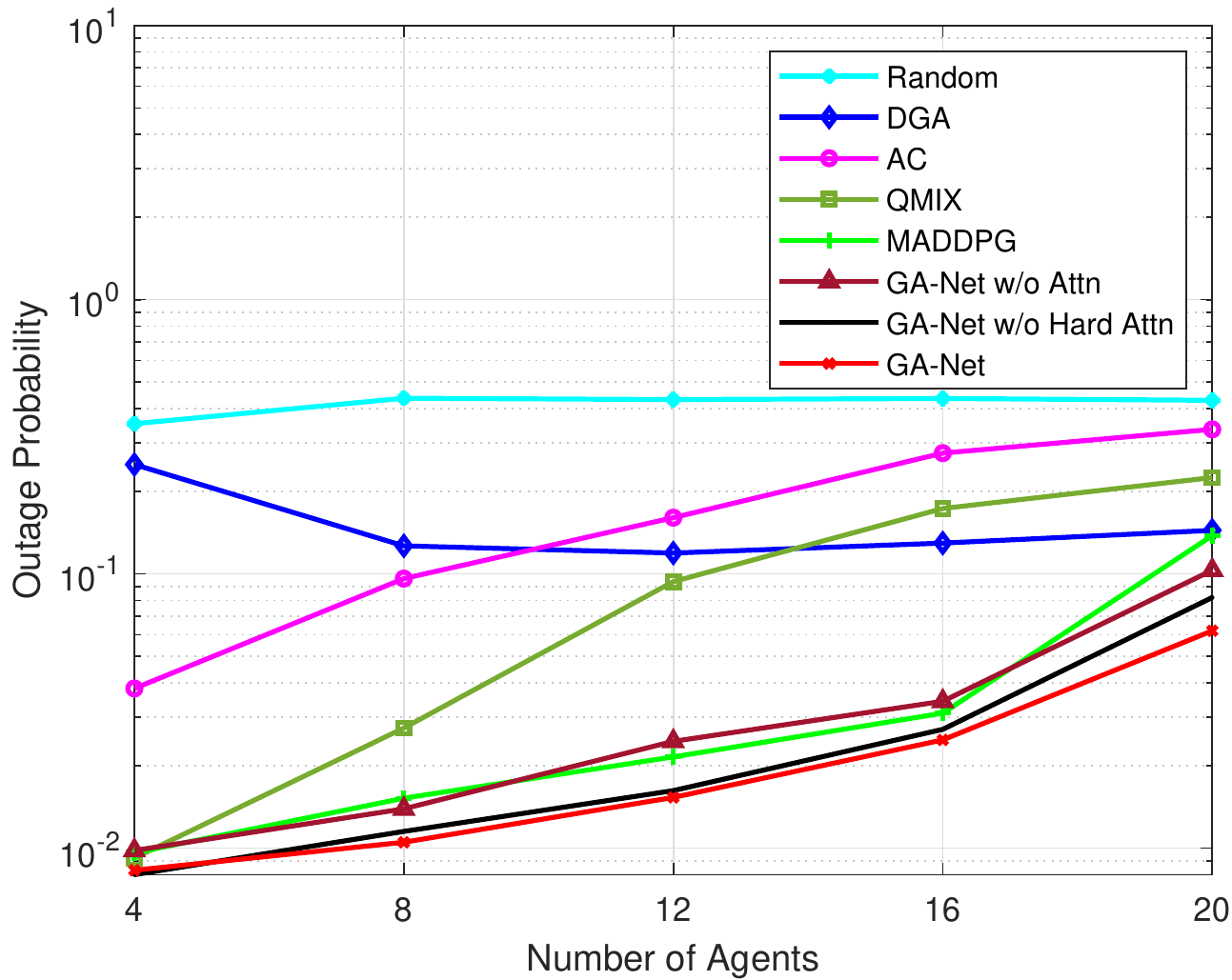}  
	\caption{Outage probabilities under different subnetwork densities with fixed channel bandwidth $100$ KHz} \label{Outage_Vs_No}
\end{figure}

In addition, we evaluate the reliability performance under different subnetwork densities. The reliability of the transmission links is decided by the outage probability of subnetworks. Fig. \ref{Outage_Vs_No} exhibits the experimental results of the outage probabilities under different subnetwork densities with fixed channel bandwidth $100$ KHz. The number of subnetworks varies from $4$ to $20$. 
We compare our approach with the Random Baseline, dynamic greedy approach and  five other DRL-based approaches (i.e., AC, MADDPG, QMIX, GA-Net w/o Hard Attn and GA-Net w/o Attn). For dynamic greedy approach, every subnetwork selects the best channel with the least RSSI without state exchange. 
When the channel number is limited, all subnetworks will choose the spare channels, which will lead to a crash with greedy selection.
It clearly can be seen from Fig. \ref{Outage_Vs_No} that the outage probabilities of seven algorithms increase as the number of subnetworks grows except the greedy approach.  This implies that a higher density leads to lower reliability because more subnetworks will lead to severer mutual interference. 
Evidently, GA-Net, GA-Net w/o Hard Attn, GA-Net w/o Attn and MADDPG have better performance than the other three algorithms because these two approaches utilize the states and actions of other subnetworks to assist in training policy during the centralized training stage, which can solve the problem of environmental instability and is conducive to better cooperation between subnetworks. Moreover, when there are 4 subnetworks in the environment, QMIX has better performance. However, as the number of subnetworks increases, the outage probabilities rises sharply.
Comparatively, our GA-Net has the lowest outage rate, especially when the number of subnetworks increases greatly, which is attributed to that GAT-Net helps subnetworks concentrate on other subnetworks with critical impact. 
As our method with GA-Net makes it easier to learn the inherent topology structural information of the environment, modeled as a graph, our approach can extract critical interference features from RSSI.
In addition, for GA-Net w/o Hard Attn algorithm, when the number of subnetworks is small, the outage probabilities have a slight increase, but when the number of subnetworks is large, the outage probabilities have a great increase. Evidently, the hard attention will be more helpful to improve the performance of 
GA-Net when there are more subnets. 
The GA-Net w/o Attn algorithm has a relatively obvious drop in outage probability under different numbers of subnetworks. Therefore, the soft attention mechanism plays a key role in the excellent performance of GA-Net.

\begin{figure}[htb]  
	\centering  
	\includegraphics[width=8.5cm]{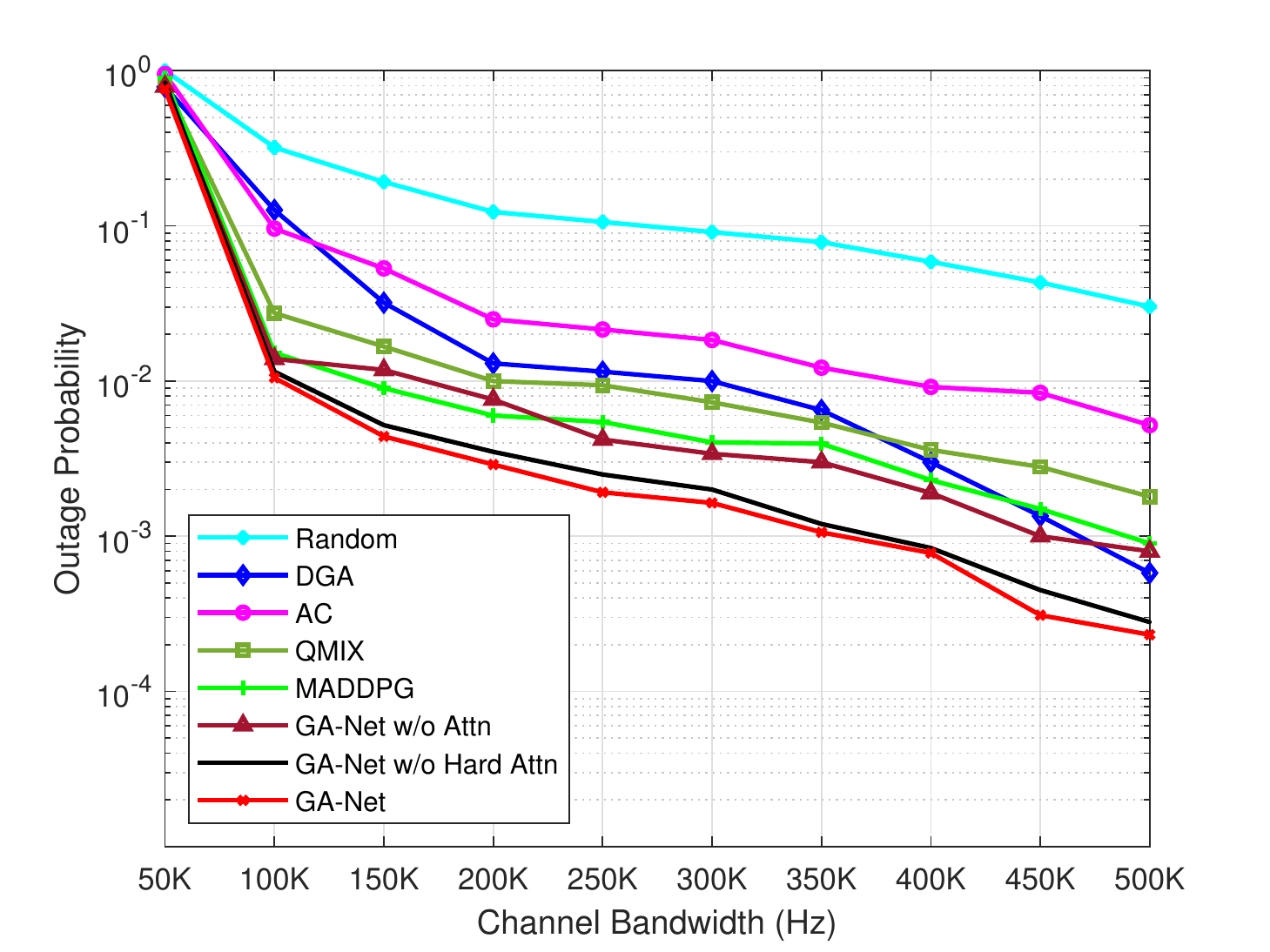}  
	\caption{Different outage probability under fixed subnetwork density and different channel bandwidth} \label{outage_Vs_bandwidth}
\end{figure}

Empirically, the reliability is an essential requirement in 5G and 6G systems. To better validate the reliability of our framework, we evaluate the outage probability of subnetworks under different bandwidth settings. 
With a fixed number of channels, for a given outage probability, the more bandwidth occupation means lower spectrum utility. Fig. \ref{outage_Vs_bandwidth} presents the different outage probabilities of all five methods with respect to different channel bandwidths.  All the compared algorithms are evaluated with $8$ subnetworks $6$ channels. The channel bandwidth varies from $50$ KHz to $500$ KHz. The simulation results show that the outage drops for all methods as the channel bandwidth grows because higher bandwidth will lead to faster transmitting speed.
For a given bandwidth, our GA-Net based algorithm achieves the least outage probability, which convinces that our approach yields the best reliability. In particular, this characteristic can reduce the total bandwidth for subnetworks, so that there are more spectrums left used for other services.  It is worth noting that the compared policy gradient algorithms (ie. MADDPPG, GAT-NET, GAT-Net w/0 Hard Attn and GAT-Net w/0 Attn) have achieved better performance than QMIX under different bandwidth.

\begin{figure}[htb]  
	\centering  
	\includegraphics[width=8.5cm]{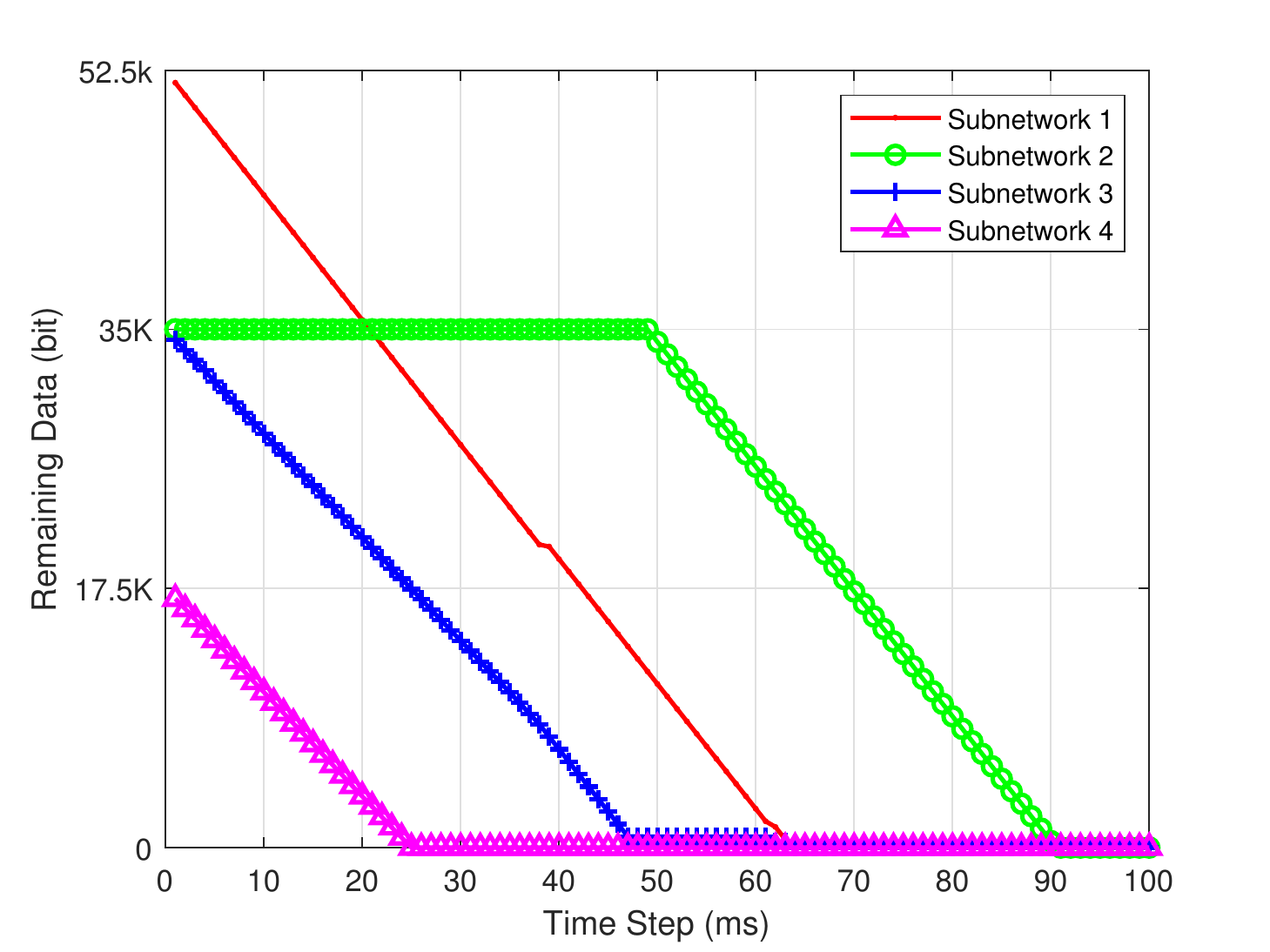}  
	\caption{The outage probability under different payload requirements} \label{Qos}
\end{figure}
Quality of service (QoS) is a mechanism that works on a network to control traffic and ensure the performance of critical applications with limited network capacity. 
For conventional algorithms, such as random baseline and DGA, the way to implement QoS requirement is to offer ultra wide channel bandwidth that can satisfy the most critical interference case. Comparatively, our GA-Net can offer flexible channel capacity with dynamic channel allocation under the constraint of bandwidth limitation. Fig. \ref{Qos} presents the experimental results of the QoS-based allocation with 4 subnetowrks and 3 channels. The scenario is the same as the one defined in Fig.6, but different subnetworks have different payload requirements, which are 17K, 34K, 34K and 51K bits, respectively. The results show that all subnetworks can transmit their target payloads in 100 ms, which proves the good ability of our GA-Net in building the network QoS policy for prioritized services, and different QoS configurations can be implemented only by GA-Net actor network update.

\subsection{Discussion}
Our proposed framework fully considers the potential relationships among mobile subnetworks. Compared with the centralized schemes, our approach does not require global knowledge, which greatly decreases both the signaling overhead and the computational burden of the base station. For other distributed methods, the `unavailable’ information like source output power and the instantaneous channel gain should be known, and can't offer QoS service.  Compared with these schemes, our approach only uses RSSI at each spectrum band as the state input to MARL, and this is more challenging to take the correct actions from the continuously changing total received power. It also can meet the flexible QoS target for different subnetworks, which will bring great benefits for communication.

\section{Conclusion} \label{conclusion}
This paper investigated the resource management problem in mobile subnetworks, and formulated it as a distributed MARL model to decrease the transmission failure rate of subnetwork. Moreover, our approach employs an attention-based graph neural network to derive the potential interference relationships among subnetworks aiming to help learn better polices and decrease computing complexity. 
The experimental results show that the subnetwork can coordinately choose its channel and power level correctly, which only needs the RSSI for each subnetwork, instead of requiring to get specific channel gain between any two agents. Furthermore, compared with the existing methods, our approach can not only significantly improve convergence speed, but also drastically reduce outage rate under the condition of different subnetwork density. In addition, the policies learned by our approach have been proven to improve the channel utilization through the cooperative selection of channel and power level.


\bibliographystyle{unsrt}
\bibliography{references}

\begin{thebibliography}{10}

\bibitem{dixon_2017}
The 5G Infrastructure~Association (5GIA).
\newblock European vision for the 6g network ecosystem.
\newblock White Paper, 2021.

\bibitem{letaief2019roadmap1}
Khaled~B Letaief, Wei Chen, Yuanming Shi, Jun Zhang, and Ying-Jun~Angela Zhang.
\newblock The roadmap to 6g: Ai empowered wireless networks.
\newblock {\em IEEE Communications Magazine}, 57(8):84--90, 2019.

\bibitem{20206G4}
G.~Berardinelli, P.~Mogensen, and R.~O. Adeogun.
\newblock 6g subnetworks for life-critical communication.
\newblock In {\em 2020 2nd 6G Wireless Summit (6G SUMMIT)}, 2020.

\bibitem{2021Radio}
Tafseer Akhtar, Christos Tselios, and Ilias Politis.
\newblock Radio resource management: approaches and implementations from 4g to
  5g and beyond.
\newblock {\em Wireless Networks}, 27(1):693--734, 2021.

\bibitem{46002287}
Anders Gjendemsjo, David Gesbert, Geir~E. Oien, and Saad~G. Kiani.
\newblock Binary power control for sum rate maximization over multiple
  interfering links.
\newblock {\em IEEE Transactions on Wireless Communications}, 7(8):3164--3173,
  2008.

\bibitem{57564896}
Qingjiang Shi, Meisam Razaviyayn, Zhi-Quan Luo, and Chen He.
\newblock An iteratively weighted mmse approach to distributed sum-utility
  maximization for a mimo interfering broadcast channel.
\newblock {\em IEEE Transactions on Signal Processing}, 59(9):4331--4340, 2011.

\bibitem{68450588}
Lingyang Song, Dusit Niyato, Zhu Han, and Ekram Hossain.
\newblock Game-theoretic resource allocation methods for device-to-device
  communication.
\newblock {\em IEEE Wireless Communications}, 21(3):136--144, 2014.

\bibitem{matching2015}
Y.~Gu, W.~Saad, M.~Bennis, M.~Debbah, and Z.~Han.
\newblock Matching theory for future wireless networks: Fundamentals and
  applications.
\newblock {\em IEEE Communications Magazine}, 53(5):52--59, 2015.

\bibitem{spectrum2017}
J.~Zhao, Y.~Liu, K.~K. Chai, A.~Nallanathan, C.~Yue, and H.~Zhu.
\newblock Spectrum allocation and power control for non-orthogonal multiple
  access in hetnets.
\newblock {\em IEEE Transactions on Wireless Communications}, 16(9):5825--5837,
  2017.

\bibitem{2017FPLinQ9}
K.~Shen and Y.~Wei.
\newblock Fplinq: A cooperative spectrum sharing strategy for device-to-device
  communications.
\newblock In {\em 2017 IEEE International Symposium on Information Theory
  (ISIT)}, 2017.

\bibitem{682474510}
Navid Naderializadeh and Amir~Salman Avestimehr.
\newblock Itlinq: A new approach for spectrum sharing in device-to-device
  communication systems.
\newblock {\em IEEE Journal on Selected Areas in Communications},
  32(6):1139--1151, 2014.

\bibitem{2015ITLinQ11}
X.~Yi and G.~Caire.
\newblock Itlinq+: An improved spectrum sharing mechanism for device-to-device
  communications.
\newblock In {\em 2015 49th Asilomar Conference on Signals, Systems and
  Computers}, 2015.

\bibitem{875530012}
Mingzhe Chen, Ursula Challita, Walid Saad, Changchuan Yin, and M{\'e}rouane
  Debbah.
\newblock Machine learning for wireless networks with artificial intelligence:
  A tutorial on neural networks.
\newblock {\em arXiv preprint arXiv:1710.02913}, 9, 2017.

\bibitem{2017Artificial13}
M.~Chen, U.~Challita, W.~Saad, C.~Yin, and M.~Debbah.
\newblock Artificial neural networks-based machine learning for wireless
  networks: A tutorial.
\newblock {\em IEEE Communications Surveys and Tutorials}, 2017.

\bibitem{2020Distributed26}
R.~Adeogun, G.~Berardinelli, I.~Rodriguez, and P.~Mogensen.
\newblock Distributed dynamic channel allocation in 6g in-x subnetworks for
  industrial automation.
\newblock In {\em 2020 IEEE Globecom Workshops (GC Wkshps)}, 2020.

\bibitem{2019Spectrum15}
L.~Liang, H.~Ye, and G.~Y. Li.
\newblock Spectrum sharing in vehicular networks based on multi-agent
  reinforcement learning.
\newblock {\em IEEE Journal on Selected Areas in Communications},
  37(10):2282--2292, 2019.

\bibitem{li2019multi}
Zheng Li and Caili Guo.
\newblock Multi-agent deep reinforcement learning based spectrum allocation for
  d2d underlay communications.
\newblock {\em IEEE Transactions on Vehicular Technology}, 69(2):1828--1840,
  2019.

\bibitem{resource2013}
Phond Phunchongharn, Ekram Hossain, and Dong~In Kim.
\newblock Resource allocation for device-to-device communications underlaying
  lte-advanced networks.
\newblock {\em IEEE Wireless Communications}, 20(4):91--100, 2013.

\bibitem{resource2018}
A.~Kose and B.~Ozbek.
\newblock Resource allocation for underlaying device-to-device communications
  using maximal independent sets and knapsack algorithm.
\newblock In {\em 2018 IEEE 29th Annual International Symposium on Personal,
  Indoor and Mobile Radio Communications (PIMRC)}, 2018.

\bibitem{learning2013}
Alessandro Checco and Douglas~J. Leith.
\newblock Learning-based constraint satisfaction with sensing restrictions.
\newblock {\em IEEE Journal of Selected Topics in Signal Processing},
  7(5):811--820, 2013.

\bibitem{9252917}
Yifei Shen, Yuanming Shi, Jun Zhang, and Khaled~B. Letaief.
\newblock Graph neural networks for scalable radio resource management:
  Architecture design and theoretical analysis.
\newblock {\em IEEE Journal on Selected Areas in Communications},
  39(1):101--115, 2021.

\bibitem{2019A20}
K.~I. Ahmed and E.~Hossain.
\newblock A deep q-learning method for downlink power allocation in multi-cell
  networks.
\newblock 2019.

\bibitem{732222617}
Hoang-Hiep Nguyen, Mikio Hasegawa, and Won-Joo Hwang.
\newblock Distributed resource allocation for d2d communications underlay
  cellular networks.
\newblock {\em IEEE Communications Letters}, 20(5):942--945, 2016.

\bibitem{2016A18}
E.~Ghadimi, F.~D. Calabrese, G.~Peters, and P.~Soldati.
\newblock A reinforcement learning approach to power control and rate
  adaptation in cellular networks.
\newblock In {\em IEEE International Conference on Communications}, 2016.

\bibitem{2017Distributed16}
F.~W. Zaki, S.~Kishk, and N.~H. Almofari.
\newblock Distributed resource allocation for d2d communication networks using
  auction.
\newblock In {\em 34 Th National Radio Science Conference}, pages 284--293,
  2017.

\bibitem{876143119}
Fan Meng, Peng Chen, and Lenan Wu.
\newblock Power allocation in multi-user cellular networks with deep q learning
  approach.
\newblock In {\em ICC 2019 - 2019 IEEE International Conference on
  Communications (ICC)}, pages 1--6, 2019.

\bibitem{879211721}
Yasar~Sinan Nasir and Dongning Guo.
\newblock Multi-agent deep reinforcement learning for dynamic power allocation
  in wireless networks.
\newblock {\em IEEE Journal on Selected Areas in Communications},
  37(10):2239--2250, 2019.

\bibitem{9565875}
Zhenjiang Shi, Jiajia Liu, Shangwei Zhang, and Nei Kato.
\newblock Multi-agent deep reinforcement learning for massive access in 5g and
  beyond ultra-dense noma system.
\newblock {\em IEEE Transactions on Wireless Communications}, pages 1--1, 2021.

\bibitem{lu2021dynamic}
Ziyang Lu, Chen Zhong, and M~Cenk Gursoy.
\newblock Dynamic channel access and power control in wireless interference
  networks via multi-agent deep reinforcement learning.
\newblock {\em IEEE Transactions on Vehicular Technology}, 2021.

\bibitem{naderializadeh2021resource}
Navid Naderializadeh, Jaroslaw~J Sydir, Meryem Simsek, and Hosein Nikopour.
\newblock Resource management in wireless networks via multi-agent deep
  reinforcement learning.
\newblock {\em IEEE Transactions on Wireless Communications}, 20(6):3507--3523,
  2021.

\bibitem{2021Distributed}
A.~Doshi and J.~G. Andrews.
\newblock Distributed deep reinforcement learning for adaptive medium access
  and modulation in shared spectrum.
\newblock {\em arXiv preprint arXiv:2109.11723}, 2021.

\bibitem{2017Multi33}
Ryan Lowe, Yi~Wu, Aviv Tamar, and Jean Harb.
\newblock Multi-agent actor-critic for mixed cooperative-competitive
  environments.
\newblock 2017.

\bibitem{tarmac2019}
Abhishek Das, Th{\'e}ophile Gervet, Joshua Romoff, Dhruv Batra, Devi Parikh,
  Mike Rabbat, and Joelle Pineau.
\newblock Tarmac: Targeted multi-agent communication.
\newblock In {\em International Conference on Machine Learning}, pages
  1538--1546. PMLR, 2019.

\bibitem{liu2020multi}
Yong Liu, Weixun Wang, Yujing Hu, Jianye Hao, Xingguo Chen, and Yang Gao.
\newblock Multi-agent game abstraction via graph attention neural network.
\newblock In {\em Proceedings of the AAAI Conference on Artificial
  Intelligence}, volume~34, pages 7211--7218, 2020.

\bibitem{2018Value31}
P.~Sunehag, G.~Lever, Audrxb, N.~Gruslys, W.~M. Czarnecki, V.~Zambaldi,
  M.~Jaderberg, M.~Lanctot, N.~Sonnerat, and J.~Z. Leibo.
\newblock Value-decomposition networks for cooperative multi-agent learning
  based on team reward.
\newblock {\em AAMAS}, 2018.

\bibitem{2018QMIX32}
T.~Rashid, M.~Samvelyan, Cs~De Witt, G.~Farquhar, J.~Foerster, and S.~Whiteson.
\newblock Qmix: Monotonic value function factorisation for deep multi-agent
  reinforcement learning.
\newblock {\em CoRR}, abs/1803.11485, 2018.

\bibitem{son2019qtran}
Kyunghwan Son, Daewoo Kim, Wan~Ju Kang, David~Earl Hostallero, and Yung Yi.
\newblock Qtran: Learning to factorize with transformation for cooperative
  multi-agent reinforcement learning.
\newblock In {\em International Conference on Machine Learning}, pages
  5887--5896. PMLR, 2019.

\bibitem{wang2020qplex}
Jianhao Wang, Zhizhou Ren, Terry Liu, Yang Yu, and Chongjie Zhang.
\newblock Qplex: Duplex dueling multi-agent q-learning.
\newblock {\em arXiv preprint arXiv:2008.01062}, 2020.

\bibitem{iqbal2021randomized}
Shariq Iqbal, Christian A~Schroeder De~Witt, Bei Peng, Wendelin B{\"o}hmer,
  Shimon Whiteson, and Fei Sha.
\newblock Randomized entity-wise factorization for multi-agent reinforcement
  learning.
\newblock In {\em International Conference on Machine Learning}, pages
  4596--4606. PMLR, 2021.

\bibitem{jiang2018graph}
Jiechuan Jiang, Chen Dun, Tiejun Huang, and Zongqing Lu.
\newblock Graph convolutional reinforcement learning.
\newblock {\em arXiv preprint arXiv:1810.09202}, 2018.

\bibitem{naderializadeh2020graph}
Navid Naderializadeh, Fan~H Hung, Sean Soleyman, and Deepak Khosla.
\newblock Graph convolutional value decomposition in multi-agent reinforcement
  learning.
\newblock {\em arXiv preprint arXiv:2010.04740}, 2020.

\bibitem{malysheva2019magnet}
Aleksandra Malysheva, Daniel Kudenko, and Aleksei Shpilman.
\newblock Magnet: Multi-agent graph network for deep multi-agent reinforcement
  learning.
\newblock In {\em 2019 XVI International Symposium" Problems of Redundancy in
  Information and Control Systems"(REDUNDANCY)}, pages 171--176. IEEE, 2019.

\bibitem{bai2021value}
Yunpeng Bai, Chen Gong, Bin Zhang, Guoliang Fan, and Xinwen Hou.
\newblock Value function factorisation with hypergraph convolution for
  cooperative multi-agent reinforcement learning.
\newblock {\em arXiv preprint arXiv:2112.06771}, 2021.

\bibitem{2018Soft34}
T.~Haarnoja, A.~Zhou, P.~Abbeel, and S.~Levine.
\newblock Soft actor-critic: Off-policy maximum entropy deep reinforcement
  learning with a stochastic actor.
\newblock 2018.

\bibitem{2016Categorical35}
E.~Jang, S.~Gu, and B.~Poole.
\newblock Categorical reparameterization with gumbel-softmax.
\newblock {\em arXiv e-prints}, 2016.

\bibitem{2017Multiagent29}
T.~Ardi, M.~Tambet, K.~Dorian, K.~Ilya, K.~Kristjan, A.~Juhan, A.~Jaan,
  V.~Raul, and C.~Y. Xia.
\newblock Multiagent cooperation and competition with deep reinforcement
  learning.
\newblock {\em Plos One}, 12(4):e0172395, 2017.

\bibitem{2017Counterfactual30}
J.~Foerster, G.~Farquhar, T.~Afouras, N.~Nardelli, and S.~Whiteson.
\newblock Counterfactual multi-agent policy gradients.
\newblock 2017.

\bibitem{velivckovic2017graph}
P.r Veli{\v{c}}kovi{\'c}, G.~Cucurull, A.~Casanova, A.~Romero, P.~Lio, and
  Y.~Bengio.
\newblock Graph attention networks.
\newblock {\em arXiv preprint arXiv:1710.10903}, 2017.

\bibitem{2014Adam22}
D.~P. Kingma and J.~Ba.
\newblock Adam: A method for stochastic optimization.
\newblock {\em arXiv e-prints}, 2014.

\bibitem{googlecolab25}
Google Colab.
\newblock Google colab.
\newblock https://colab.research.google.com.

\bibitem{2015Continuous}
Timothy~P. Lillicrap, Jonathan~J. Hunt, Alexander Pritzel, Nicolas Heess, Tom
  Erez, Yuval Tassa, David Silver, and Daan Wierstra.
\newblock Continuous control with deep reinforcement learning.
\newblock {\em Computer ence}, 2015.

\end{thebibliography}

\end{document}